\documentclass{article}
\usepackage{graphicx} % Required for inserting images
\usepackage{subcaption}
\usepackage{float}

\usepackage{comment}
\usepackage{url}
\usepackage[hidelinks]{hyperref}

\title{Cross-Platform Digital Discourse Analysis of Iran: Topics, Sentiment, Polarization, and Event Validation on Telegram and Reddit}
\author{
Despoina Antonakaki$^{1,2}$,
Sotiris Ioannidis$^{2}$ \\
\\
$^{1}$Institute of Computer Science, Foundation for Research and Technology,\\ Vassilika Vouton, Heraklion, Crete, Greece \\
$^{2}$Technical University of Crete, University Campus, \\Akrotiri, Chania, Greece \\
} 
\begin{document}

\maketitle

\begin{abstract}
 
        We analyze Iran-related discourse across two structurally different platforms: Telegram (7,567 messages from international news channels) and Reddit (23,909 posts and comments from Iran-focused and global communities). Using a single reproducible pipeline, we apply NMF topic modeling over TF--IDF features, VADER sentiment scoring, and a keyword-bundle escalation index capturing military, nuclear, and diplomatic narratives.        To assess whether discourse dynamics track offline developments, we compare escalation time series with external protest and geopolitical event timelines using same-day and lagged correlation analysis. Same-day correlations are weak, but the strongest relationships occur at non-zero lags, consistent with anticipatory or reactive framing rather than instantaneous mirroring. Finally, using a separate real-time collection (February 2026), we observe synchronized increases in escalation-related narratives that coincide with documented geopolitical developments.
        Overall, the results show systematic cross-platform differences in narrative structure and tone, and provide quantitative evidence that online escalation signals can align with real-world developments with measurable temporal offsets.
         
\end{abstract}

\section{Introduction}

Online platforms play a central role in shaping public discourse around geopolitical events and state-related narratives. In the context of Iran, digital communication environments such as Telegram and Reddit have become key arenas for information dissemination, political debate, and narrative contestation. Telegram primarily supports broadcast-style news distribution, while Reddit enables threaded, community-driven discussion. These structural differences make cross-platform analysis particularly valuable for understanding how geopolitical narratives emerge, evolve, and polarize across heterogeneous audiences.

This paper presents a comprehensive cross-platform analysis of Iran-related discourse on Telegram and Reddit. Using a unified dataset collected directly from both platforms, we examine message volume, topic structure, sentiment dynamics, polarization patterns, and temporal narrative evolution. We introduce a composite escalation index based on military, nuclear, diplomatic, and escalation-related keyword bundles, enabling systematic quantification of narrative escalation over time. To evaluate whether online discourse reflects real-world developments, we validate escalation signals against external protest and geopolitical event timelines using lagged correlation analysis.

Unlike prior studies that focus on individual platforms or apply topic modeling and sentiment analysis in isolation, this work integrates cross-platform discourse modeling, polarization measurement, escalation detection, and event validation within a single reproducible analytical framework. In addition to retrospective analysis, we demonstrate real-time escalation detection using newly collected data, bridging descriptive discourse analysis with operational monitoring of emerging geopolitical developments.
\section{Research Questions and Contributions}
\label{sec:rq_contrib}
The study is guided by the following research questions:

\begin{itemize}

\item How does Iran-related discourse differ structurally between Telegram and Reddit?
\item How do sentiment dynamics differ across platforms, and what do these differences reveal about narrative tone and polarization?
\item Can escalation-related discourse be quantitatively detected using keyword-based indicators aggregated into a composite index?
\item Do online escalation signals correspond to documented real-world geopolitical events, and do they exhibit anticipatory or reactive temporal patterns?

\end{itemize}

\vspace{0.5em}
To answer these questions, this paper makes the following contributions:

\begin{itemize}

\item \textbf{Large-Scale Cross-Platform Dataset.}
We construct and analyze a multi-year dataset comprising 7,567 Telegram messages and 23,909 Reddit items, enabling systematic comparison between broadcast-style and community-driven discourse environments.

\item \textbf{Unified Cross-Platform Analytical Framework.}
We introduce an integrated pipeline combining TF--IDF representations, Non-negative Matrix Factorization (NMF) topic modeling, VADER sentiment analysis, polarization assessment, and keyword-based escalation detection within a consistent analytical framework.

\item \textbf{Composite Escalation Index.}
We propose a normalized escalation index derived from military, nuclear, diplomatic, and escalation-related keyword bundles, providing an interpretable quantitative measure of narrative escalation intensity.

\item \textbf{Event Validation with Lag Analysis.}
We validate escalation signals using external protest and geopolitical event datasets, performing same-day and lagged correlation analysis to assess reactive and anticipatory discourse dynamics.

\item \textbf{Real-Time Escalation Demonstration.}
Using newly collected data, we demonstrate that the proposed framework detects emerging geopolitical escalation in near real time, showing alignment between discourse amplification and documented events.

\end{itemize}

Collectively, these contributions establish a reproducible and interpretable cross-platform framework for detecting, characterizing, and validating geopolitical narrative escalation in online environments.

\section{Dataset}
\label{sec:dataset}
\subsection{Data Collection}

Data were collected from Telegram and Reddit using custom-built Python crawlers designed for incremental and reproducible data acquisition. 
Telegram data were retrieved using the Telethon library, monitoring a fixed set of international news and media channels. 
Messages were stored in JSON Lines (JSONL) format, with one message per line, allowing for safe appending and recovery in case of interruption.
Reddit data were collected using Reddit's public JSON endpoints. Posts and their complete comment trees were retrieved from a set of Iran-focused subreddits as well as broader political and world news subreddits. 
Each Reddit post and comment was stored as a separate JSONL entry, enabling fine-grained analysis of both content types.
 
The dataset spans multiple years and platforms. Telegram data cover the period from November 10, 2017 to February 11, 2026, comprising 7,567 messages obtained from monitored international news channels. 
Reddit data cover the period from January 12, 2025 to February 11, 2026, comprising 23,909 posts and comments collected from Iran-related and global news subreddits. 
All data were stored locally for offline analysis, and duplicate entries were removed based on platform-specific identifiers.

\subsection{Dataset Statistics}

Table~\ref{tab:dataset_stats} summarizes the composition of the dataset across platforms.

\begin{table}[h]
\centering
\begin{tabular}{lrr}
\hline
\textbf{Platform} & \textbf{Content Type} & \textbf{Count} \\
\hline
Telegram & Messages & 7,567 \\
\hline
Reddit & Posts & 1,996 \\
Reddit & Comments & 21,913 \\
Reddit & Total Items & 23,909 \\
\hline
\textbf{All Platforms} & \textbf{Total Items} & \textbf{31,476} \\
\hline
\end{tabular}
\caption{Summary of the collected Iran-related dataset.}
\label{tab:dataset_stats}
\end{table}
The Telegram portion of the dataset primarily reflects broadcast-style communication from established media organizations, whereas Reddit data capture interactive discussions among users with diverse political orientations. This complementary structure allows for cross-platform comparison of narrative framing, sentiment, and polarization in Iran-related discourse.

\section{Methodology}
\label{meth}

This section describes the analytical pipeline used to process and analyze Iran-related discourse on Telegram and Reddit. All analyses were implemented using custom Python scripts operating on locally stored JSONL datasets to ensure reproducibility and transparency. The methodology integrates text preprocessing, topic modeling, narrative framing analysis, sentiment analysis, polarization measurement, and temporal escalation detection.

\paragraph{Data Processing and Representation}

All messages were preprocessed prior to analysis. Preprocessing steps included removal of URLs, normalization of whitespace, and filtering of empty or malformed entries. Each message was associated with platform metadata, source identifiers, and timestamps, enabling temporal, platform-specific, and cross-platform analysis.
Text data were transformed into numerical representations using Term Frequency–Inverse Document Frequency (TF--IDF). This representation captures the importance of words relative to their frequency across the corpus while reducing the influence of common or uninformative terms. TF--IDF representations were used as input for topic modeling and narrative structure analysis.
Duplicate entries were removed using platform-specific identifiers, ensuring that each message or comment was represented only once. All processed data were stored locally to support reproducibility and independent verification.

\paragraph{Topic and Narrative Analysis}

Latent thematic structure was identified using Non-negative Matrix Factorization (NMF) applied to TF--IDF representations. NMF decomposes the document-term matrix into interpretable topic components, enabling identification of coherent topics without requiring predefined labels.
Each topic was characterized by its highest-weight keywords and associated messages. Topic prevalence was computed by aggregating topic assignments across time and platforms, enabling identification of dominant narratives and temporal topic evolution.
Narrative framing was further examined using keyword frequency analysis and lexicon-based bundle detection. Predefined keyword bundles representing geopolitical themes such as military activity, nuclear policy, diplomatic engagement, and escalation-related terminology were used to identify narrative trends. Keyword activity was aggregated daily to produce temporal indicators of narrative emphasis and escalation-related discourse.

\paragraph{Sentiment Analysis}
\label{subsec:sentiment}

To quantify narrative tone, we applied lexicon-based sentiment analysis using the VADER \cite{hutto2014vader} (Valence Aware Dictionary and sEntiment Reasoner) model. VADER is specifically designed for social media text and produces a compound sentiment score in the range $[-1,1]$, representing negative to positive sentiment polarity.
Each Telegram message and Reddit item (including both posts and comments) was independently scored after preprocessing. Daily aggregate sentiment values were computed separately for each platform by averaging message-level sentiment scores. To ensure statistical robustness and reduce noise from low-volume periods, only days with at least 10 messages were included in temporal sentiment analysis.
To identify sustained sentiment trends while reducing short-term volatility, a 14-day rolling mean was applied to daily sentiment values. This smoothing approach enables clearer identification of gradual shifts in narrative tone and emotional framing.
In addition to temporal trends, sentiment score distributions were analyzed per platform to characterize structural differences in emotional framing between Telegram and Reddit discourse.

\paragraph{Polarization and Divergence Measurement}

Polarization was assessed by comparing sentiment distributions across platforms and sources. Differences in sentiment structure provide insight into variations in narrative framing and emotional tone between broadcast-oriented and community-driven discourse environments.
Distributional comparisons enable identification of systematic differences in narrative tone across platforms, revealing whether discourse is consistently neutral, polarized, or escalation-oriented. These differences provide insight into the structural role of each platform in shaping geopolitical narratives.
\paragraph{Temporal and Escalation Analysis}

Temporal dynamics were analyzed by aggregating message counts, keyword frequencies, and sentiment indicators on a daily basis. Message volume serves as a proxy for attention intensity, while keyword bundle frequency provides a direct indicator of escalation-related narrative activity.
Escalation signals were identified using predefined keyword bundles representing military, nuclear, diplomatic, and escalation-related terminology. Daily keyword frequencies were normalized relative to total message volume to control for baseline activity fluctuations.
A composite escalation index was constructed by aggregating normalized keyword bundle frequencies across escalation-related categories. This index provides a unified quantitative measure of escalation-related discourse intensity and enables direct comparison across platforms and time periods.
Temporal alignment between escalation indicators and geopolitical developments was examined using time-series analysis and visualization, enabling identification of escalation periods and narrative amplification patterns.

\paragraph{Event Correlation and Entity Network Analysis}

To evaluate the relationship between online discourse and real-world geopolitical developments, escalation indicators were compared with external event timelines. Event signals were represented as daily time series corresponding to documented geopolitical developments.
Correlation analysis was performed to evaluate temporal relationships between escalation indicators and event signals, enabling assessment of whether narrative escalation precedes, coincides with, or follows real-world developments.
In addition, named entity extraction was used to identify geopolitical actors, locations, and organizations mentioned in messages. Entity co-occurrence networks were constructed by connecting entities appearing within the same message. These networks provide structural insight into relationships among geopolitical actors and help contextualize escalation-related narratives.
All intermediate results, including processed datasets, sentiment scores, escalation indicators, and visualization outputs, were stored to ensure reproducibility and enable independent validation of analytical results.
\section{Results and Validation}

\subsection{Message Volume and Source Activity}

\begin{figure}[H]
\centering

\begin{subfigure}{0.48\linewidth}
\centering
\includegraphics[width=\linewidth]{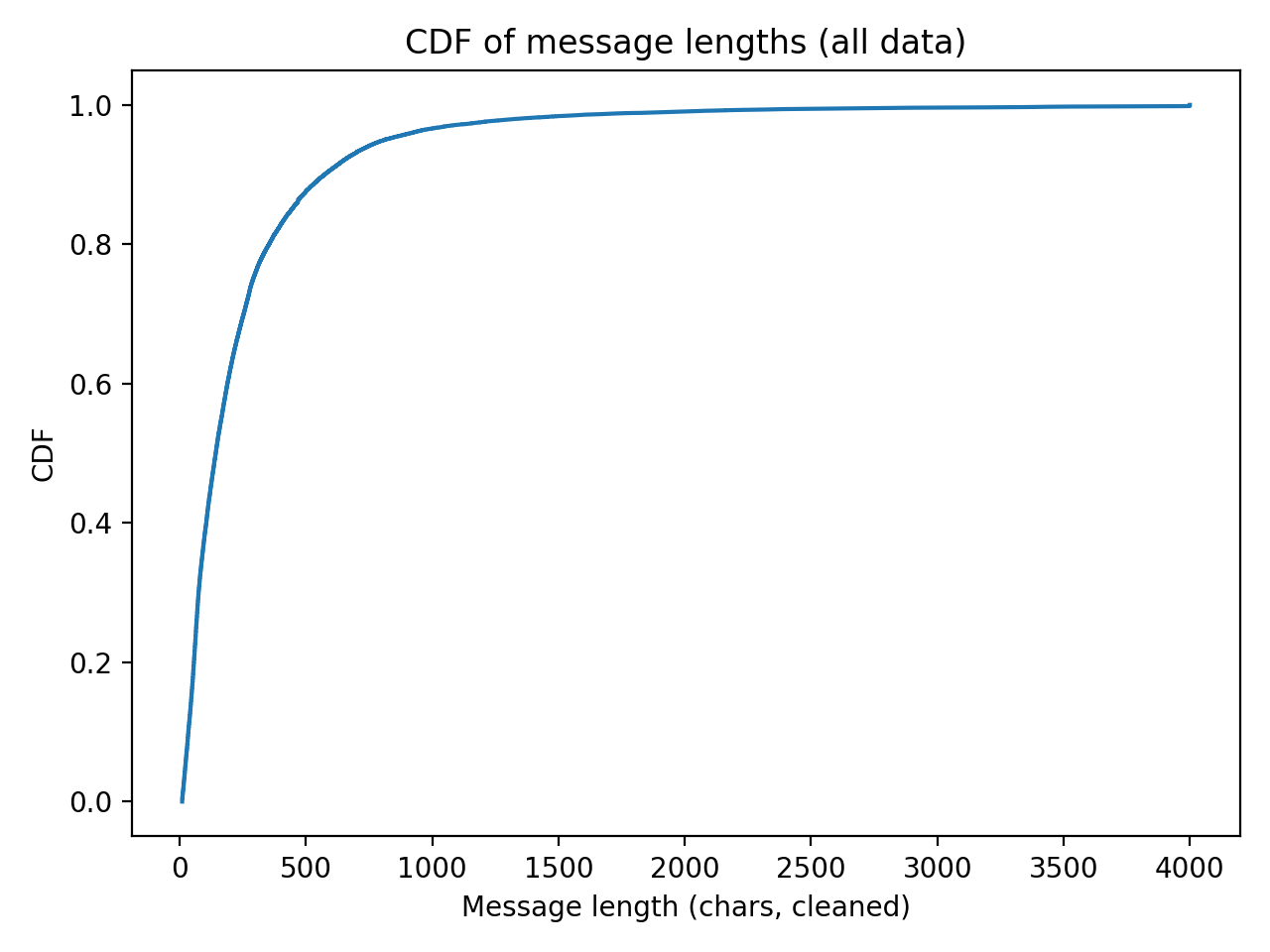}
\caption{CDF of message lengths (all data).}
\label{fig:cdf_msg_len}
\end{subfigure}
\hfill
\begin{subfigure}{0.48\linewidth}
\centering
\includegraphics[width=\linewidth]{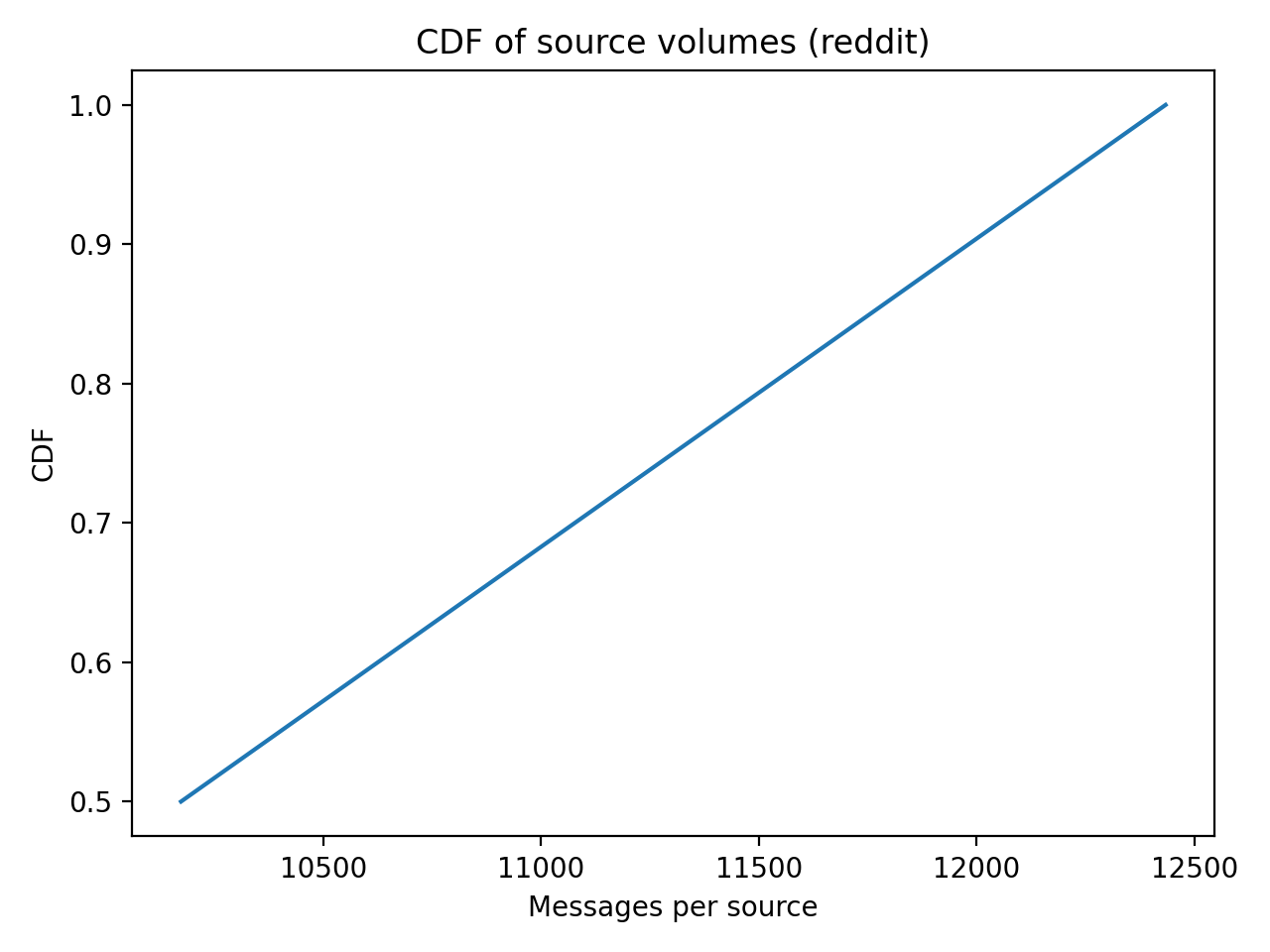}
\caption{CDF of source volumes (Reddit and Telegram).}
\label{fig:cdf_sources}
\end{subfigure}

\caption{Distributional properties of the dataset. Left: cumulative distribution of message lengths across all platforms. Right: cumulative distribution of message volumes per source, illustrating strong concentration of activity.}
\label{fig:volume_cdfs}
\end{figure}

% 1.volumne.py 
Figure~\ref{fig:volume_cdfs} summarizes distributional properties of the dataset. Figure~\ref{fig:cdf_msg_len} (left) shows the cumulative distribution function (CDF) of message lengths across the entire dataset. The distribution is highly right-skewed, indicating that the majority of messages are relatively short. Approximately 80\% of messages contain fewer than 500 characters after preprocessing, while only a small fraction exceed 1,000 characters. This pattern reflects the predominance of short-form updates and commentary, particularly on Telegram, alongside longer discussion posts on Reddit.

Figure~\ref{fig:cdf_sources} (right) presents the CDFs of message volumes per source for Reddit and Telegram. The distributions reveal a strong concentration of activity, indicating that a limited number of subreddits and Telegram channels dominate Iran-related content production on their respective platforms.

\begin{figure}[h]
\centering
\includegraphics[width=0.6\linewidth]{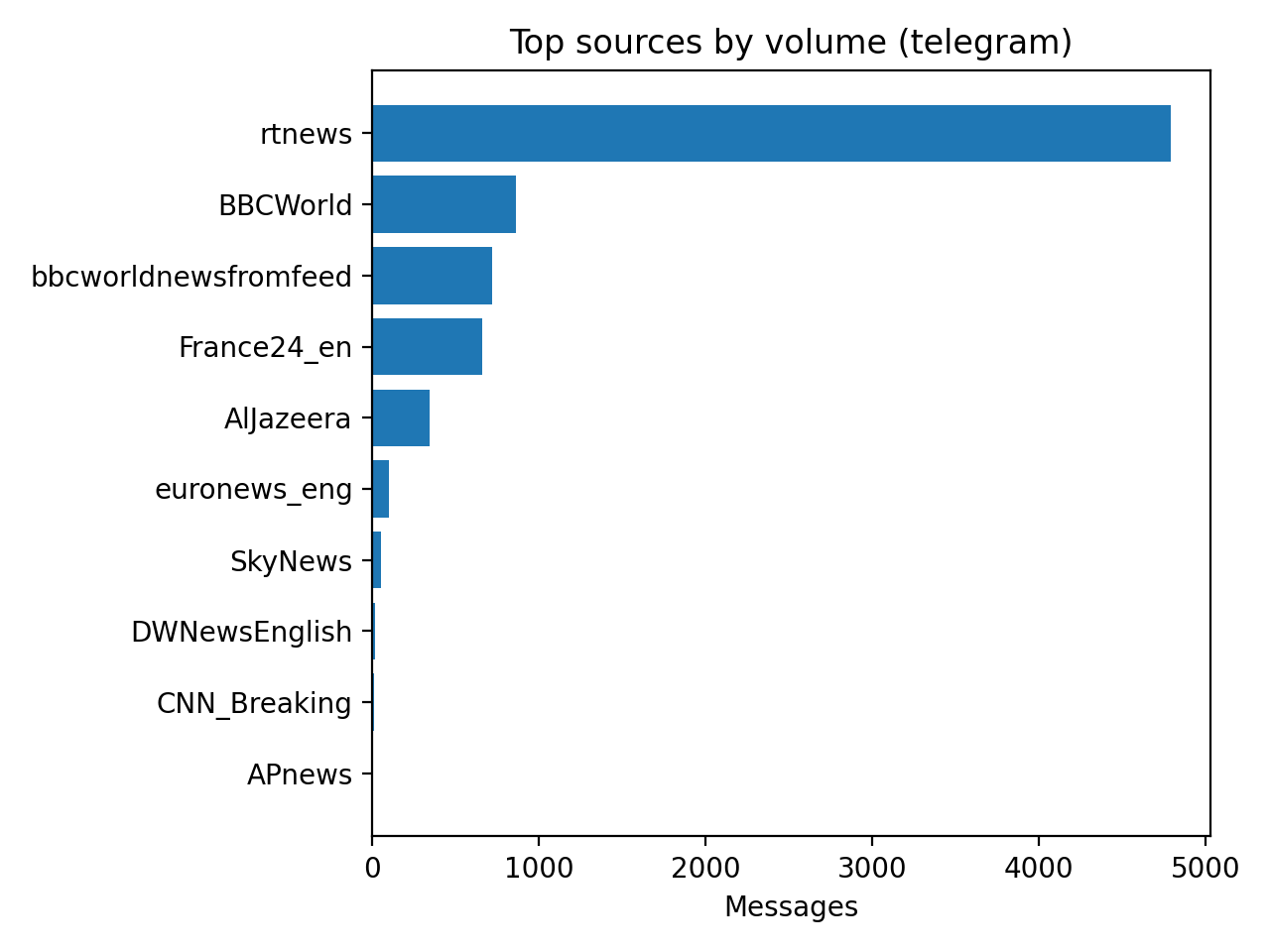}
\caption{Top Telegram channels by message volume. A small number of international news channels dominate Iran-related content production on Telegram.}
\label{fig:top_sources_telegram}
\end{figure}

Figure~\ref{fig:top_sources_telegram} further illustrates this concentration on Telegram, showing that a small number of international news channels account for a disproportionate share of Iran-related message production.
On Reddit, Iran-focused subreddits contribute the majority of posts and comments, highlighting the central role of a small number of communities in shaping Iran-related discussions. On Telegram, international news channels dominate content production, with a single channel contributing a disproportionately large number of messages compared to others. This divergence reflects structural differences between the platforms, with Reddit favoring community-centric discussion and Telegram emphasizing broadcast-style information dissemination.

\subsection{Topic Analysis}
\label{subsec:topic_analysis}

\begin{figure}[H]
    \centering
    \includegraphics[width=0.95\linewidth]{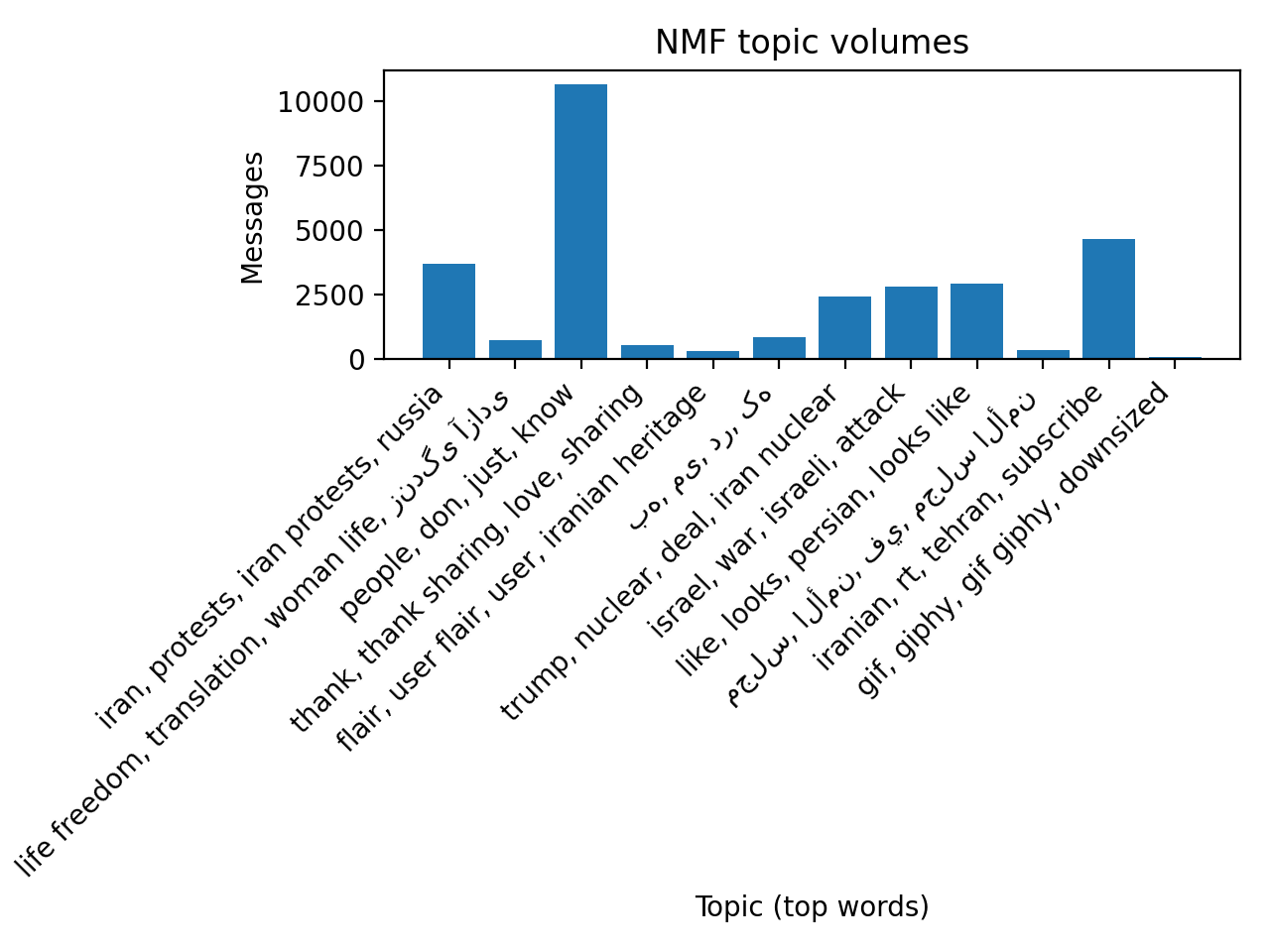}
    \caption{Message volumes per topic obtained using Non-negative Matrix Factorization (NMF). Topics are labeled using their most representative keywords. Several topics contain non-English (Persian) terms reflecting original-language discourse in the dataset.}
    \label{fig:nmf_topics_volume}
\end{figure}

Figure~\ref{fig:nmf_topics_volume} shows the distribution of message volumes across topics identified using Non-negative Matrix Factorization (NMF). The results reveal a highly uneven topic distribution, with a small number of dominant topics accounting for a substantial proportion of the total messages. These high-volume topics primarily relate to protests, international conflict, nuclear negotiations, and geopolitical actors, while lower-volume topics capture more specialized or platform-specific discussions.

Persian keywords appearing in topic labels include \emph{zan} (“woman”), \emph{zendegi} (“life”), and \emph{azadi} (“freedom”). These terms frequently co-occur in protest-related discourse and were retained in transliterated form to ensure LaTeX compatibility. The presence of both English and Persian keywords among the topic labels reflects the multilingual nature of Iran-related online discourse, particularly on Reddit.

\subsubsection{Topic Similarity Networks}
\label{subsec:topic_similarity_networks}

\begin{figure}[H]
\centering
\begin{subfigure}{0.49\linewidth}
    \centering
    \includegraphics[width=\linewidth]{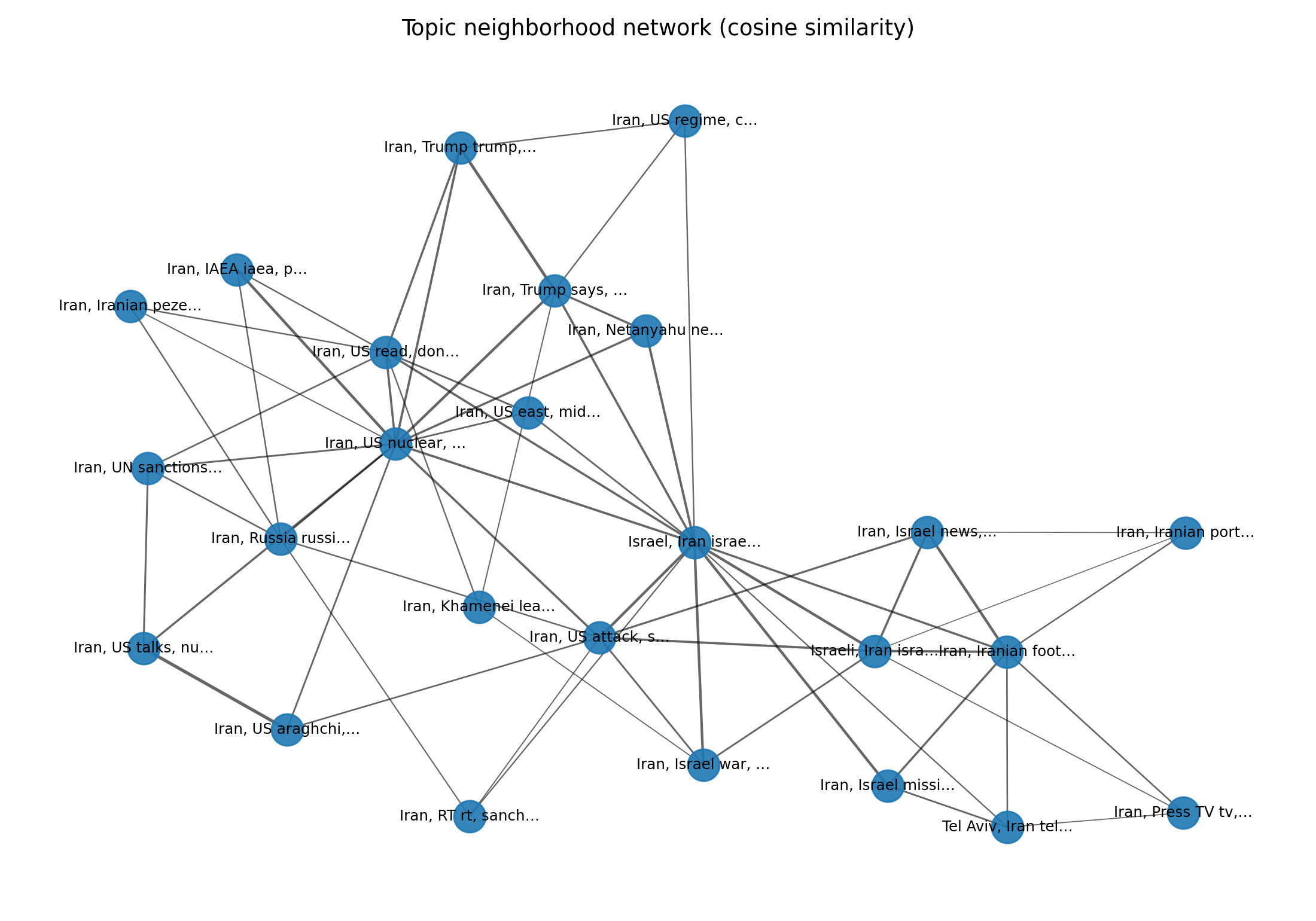}
    \caption{Telegram topic neighborhood network.}
    \label{fig:topic_network_tg}
\end{subfigure}
\hfill
\begin{subfigure}{0.49\linewidth}
    \centering
    \includegraphics[width=\linewidth]{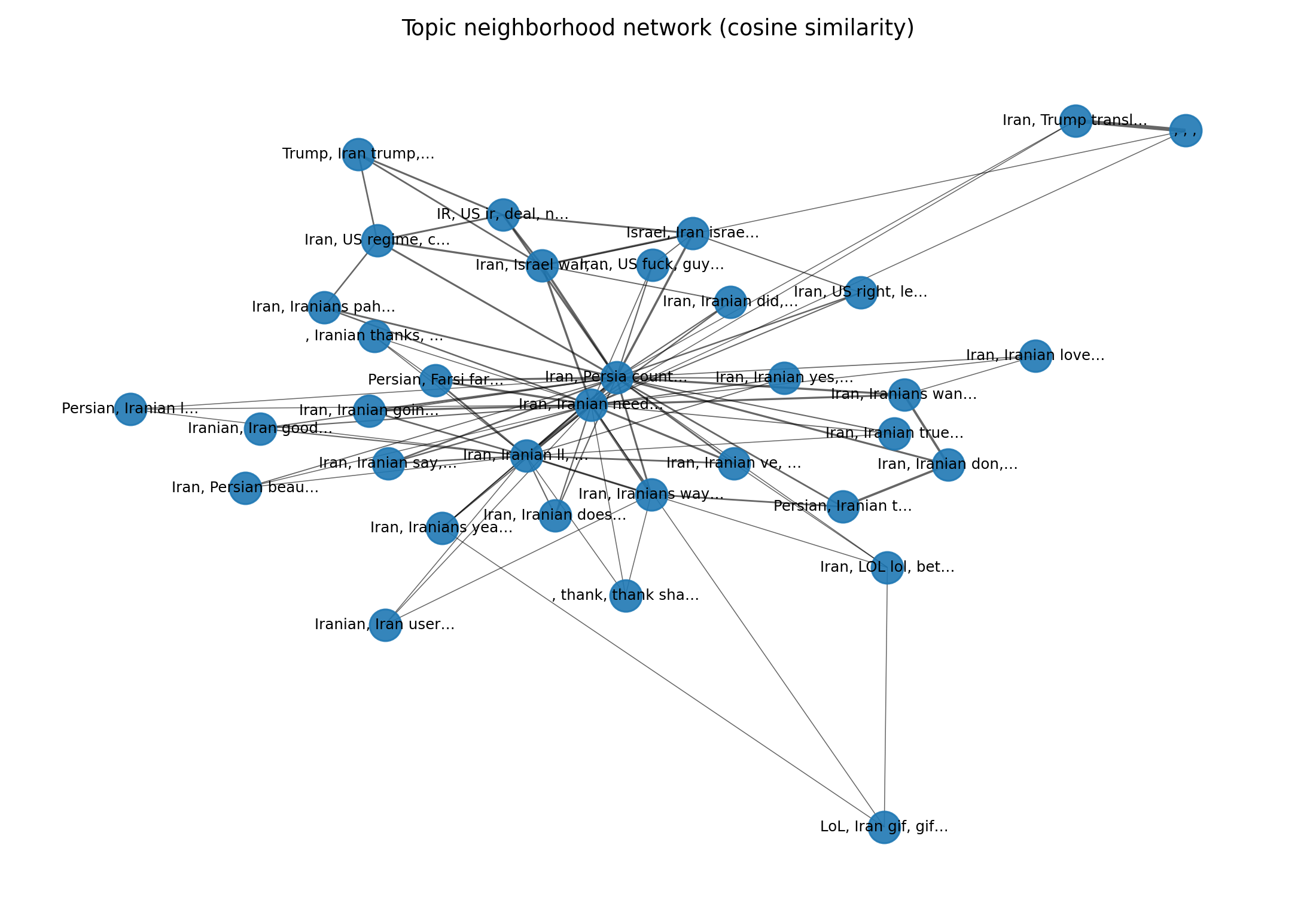}
    \caption{Reddit topic neighborhood network.}
    \label{fig:topic_network_rd}
\end{subfigure}
\caption{Topic neighborhood networks computed from pairwise cosine similarity between topic representations. Each node corresponds to one NMF topic (labeled by its most representative keywords). Edges connect each topic to its nearest neighbors under cosine similarity; thicker edges indicate higher similarity. Dense regions indicate overlapping vocabulary and closely related narrative themes, while isolated nodes indicate more distinct topics.}
\label{fig:topic_similarity_networks}
\end{figure}

Beyond topic prevalence, we examine lexical relatedness among topics using topic neighborhood networks.
Figure~\ref{fig:topic_similarity_networks} visualizes cosine similarity between topic representations, where each node is an NMF topic and edges connect each topic to its nearest neighbors.
Telegram (Figure~\ref{fig:topic_network_tg}) forms compact clusters around recurring geopolitical frames and news-driven vocabulary, consistent with broadcast-style reporting and repeated narrative templates.
Reddit (Figure~\ref{fig:topic_network_rd}) exhibits stronger overlap among conversational topics, reflecting shared community jargon and cross-topic mixing typical of participatory discussions.
Isolated nodes in both networks correspond to more specialized themes with limited vocabulary overlap.

\paragraph{Edge Interpretation}

As an illustrative example, the semantic edge between Topic 7.0 and Topic 18.0 (combined weight = 0.63) reflects sustained overlap during the period 12 June 2025 – 8 February 2026. Shared entities include Iran, Iranian, Persian, and Germany. Representative messages from these topics reference cultural heritage, diaspora identity, naming practices, and political commentary. This indicates that topic similarity arises from overlapping discourse centered on Iranian identity and transnational sociopolitical themes.

A second illustrative interconnection is observed between Topic 12.0 and Topic 18.0 (combined similarity weight = 0.59). 
The detected co-activity window (June 2025 – February 2026) indicates sustained overlap rather than a short-lived event-driven spike. 

Unlike the predominantly cultural coupling observed for T7.0–T18.0, this edge reflects the convergence of everyday domestic discourse with ideological and identity-centered narratives. Topic 12.0 includes discussions of Iranian economic life (currency usage, consumer goods, automobiles), historical memory (the Constitutional Revolution and the 1909 events in Tehran), and travel-related commentary. Topic 18.0, by contrast, contains regime critique, ideological positioning, and emotionally charged identity framing.

The shared entities — \emph{Iran}, \emph{Iranian}, \emph{Tehran}, \emph{Persian}, and \emph{Pakistani} — indicate that the overlap is structured around national identity and geopolitical positioning rather than a single discrete political event. This coupling suggests that discussions of daily life, historical memory, and national heritage become discursively intertwined with contemporary political narratives and regime evaluation.

In contrast to escalation-driven clusters (e.g., military or nuclear topics), this interconnection represents a structurally persistent identity-discourse bridge, where socio-economic concerns, historical symbolism, and ideological commentary co-evolve within the same temporal window.
\begin{figure}[H]
    \centering
    \includegraphics[width=0.90\linewidth]{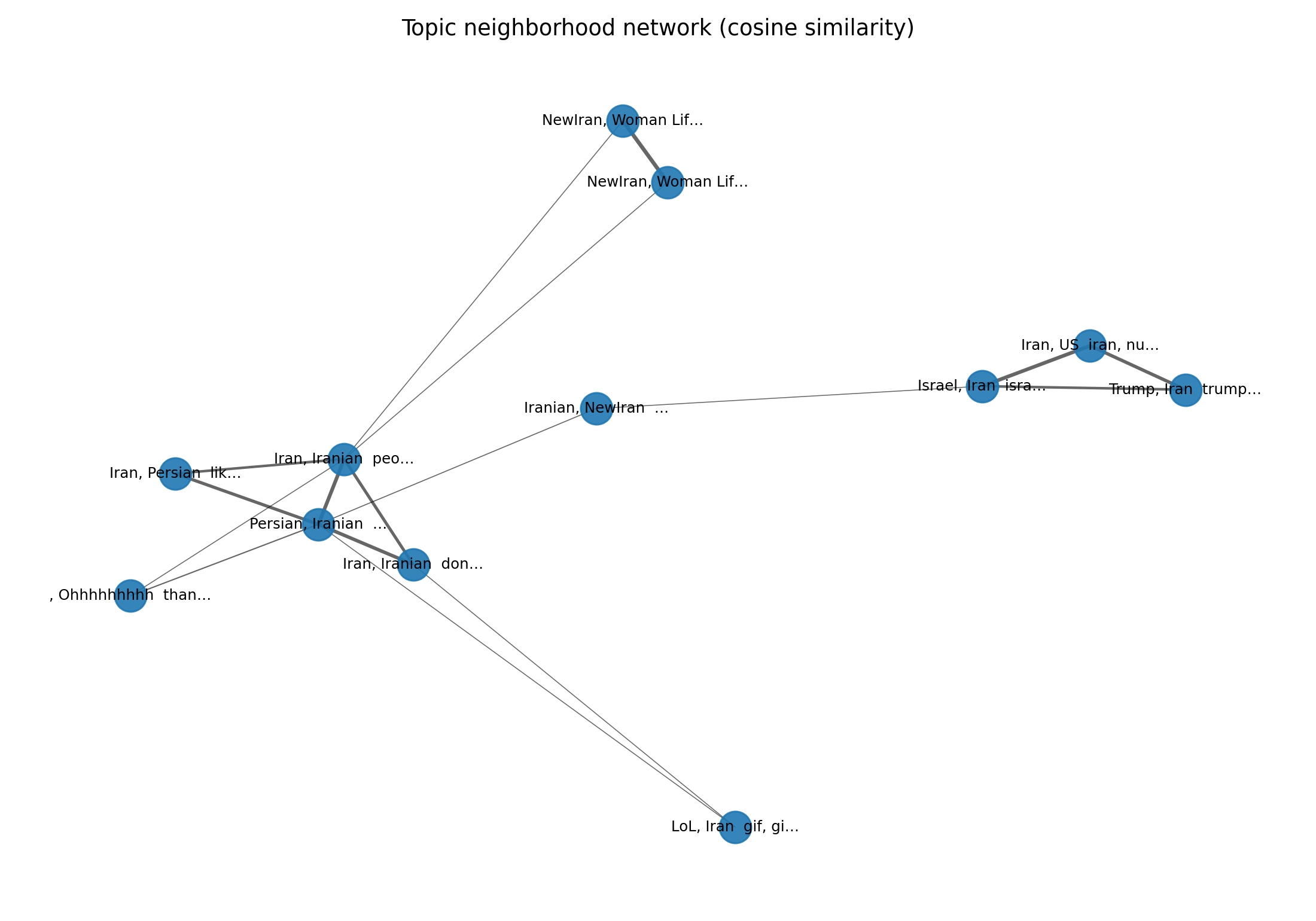}
    \caption{Topic neighborhood network computed on the combined Telegram+Reddit corpus. Nodes represent topics and edges denote nearest-neighbor cosine similarity between topic representations (thicker edges indicate higher similarity).}
    \label{fig:topic_network_all}
\end{figure}

\subsection{Sentiment Dynamics Across Platforms}
\begin{figure}[htbp]
    \centering
    \includegraphics[width=\linewidth]{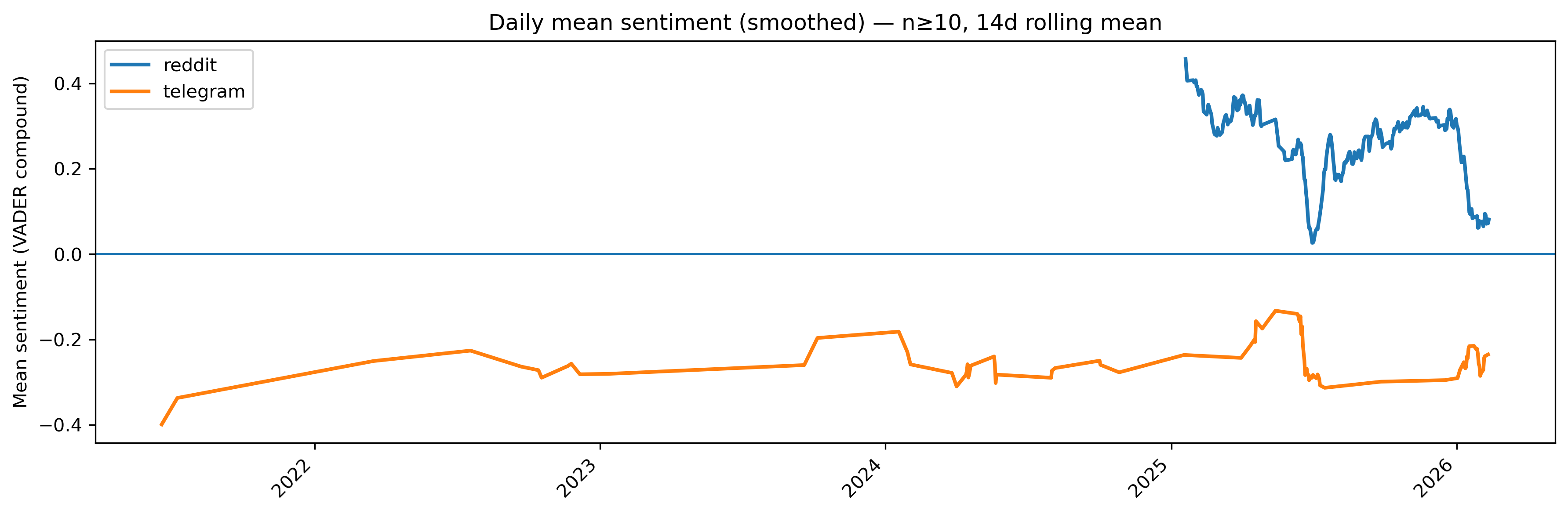}
    \caption{
    Temporal evolution of smoothed daily sentiment across Telegram and Reddit.
    Values represent 14-day rolling mean of daily VADER compound sentiment scores.
    Only days with at least 10 messages are included to ensure statistical reliability.
    Telegram exhibits persistently negative average sentiment, whereas Reddit remains predominantly neutral-to-positive with a declining trend in the final period.    }
    \label{fig:sentiment_smoothed}
\end{figure}

\begin{figure}[htbp]
    \centering
    \includegraphics[width=0.7\linewidth]{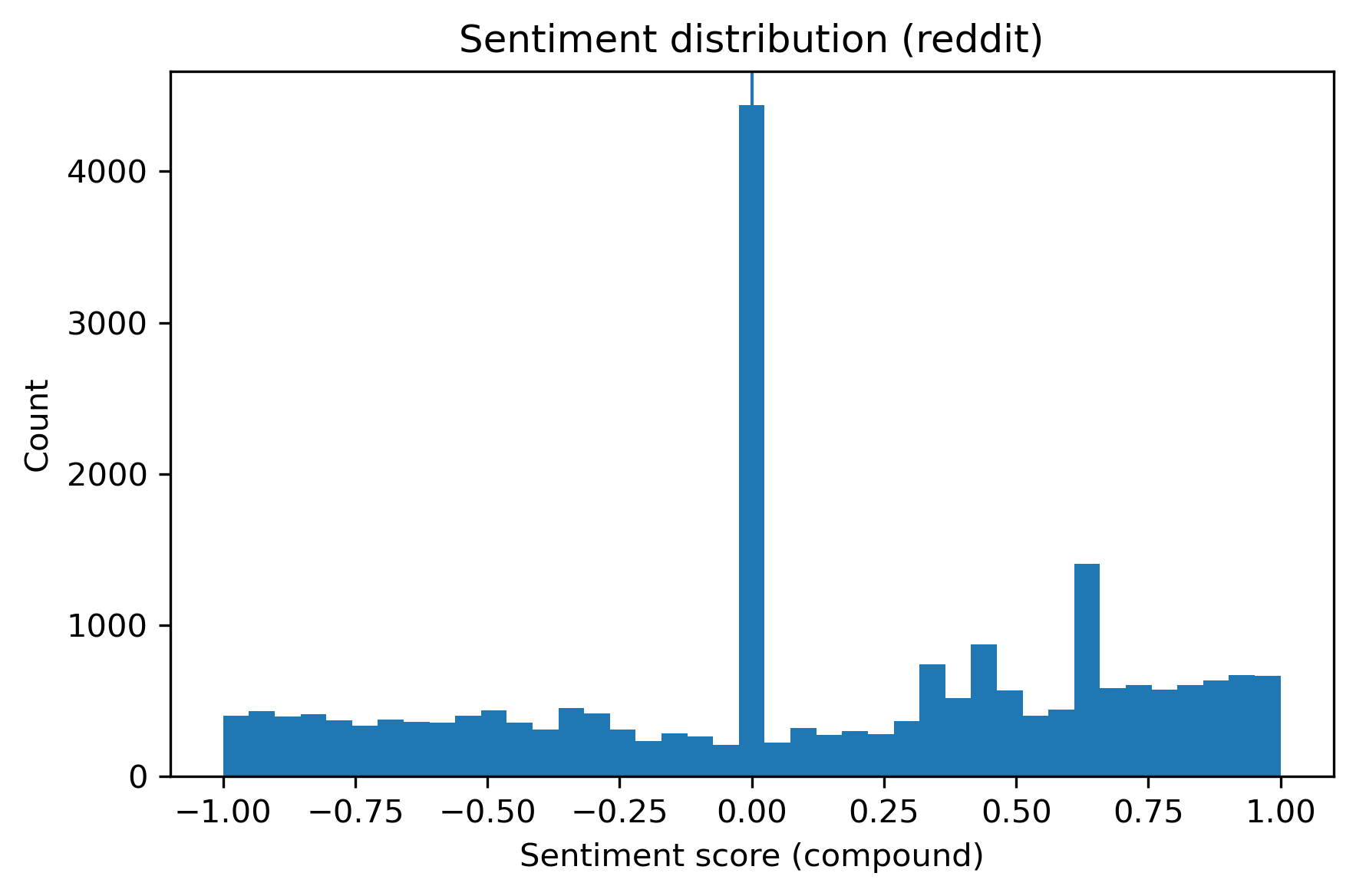}
    \caption{
    Distribution of sentiment scores for Reddit messages.
    The distribution is relatively balanced, with a substantial proportion of neutral and positive sentiment, indicating mixed narrative framing.
    }
    \label{fig:sentiment_distribution_reddit}
    \end{figure}

\begin{figure}[htbp]
        \centering
        \includegraphics[width=0.7\linewidth]{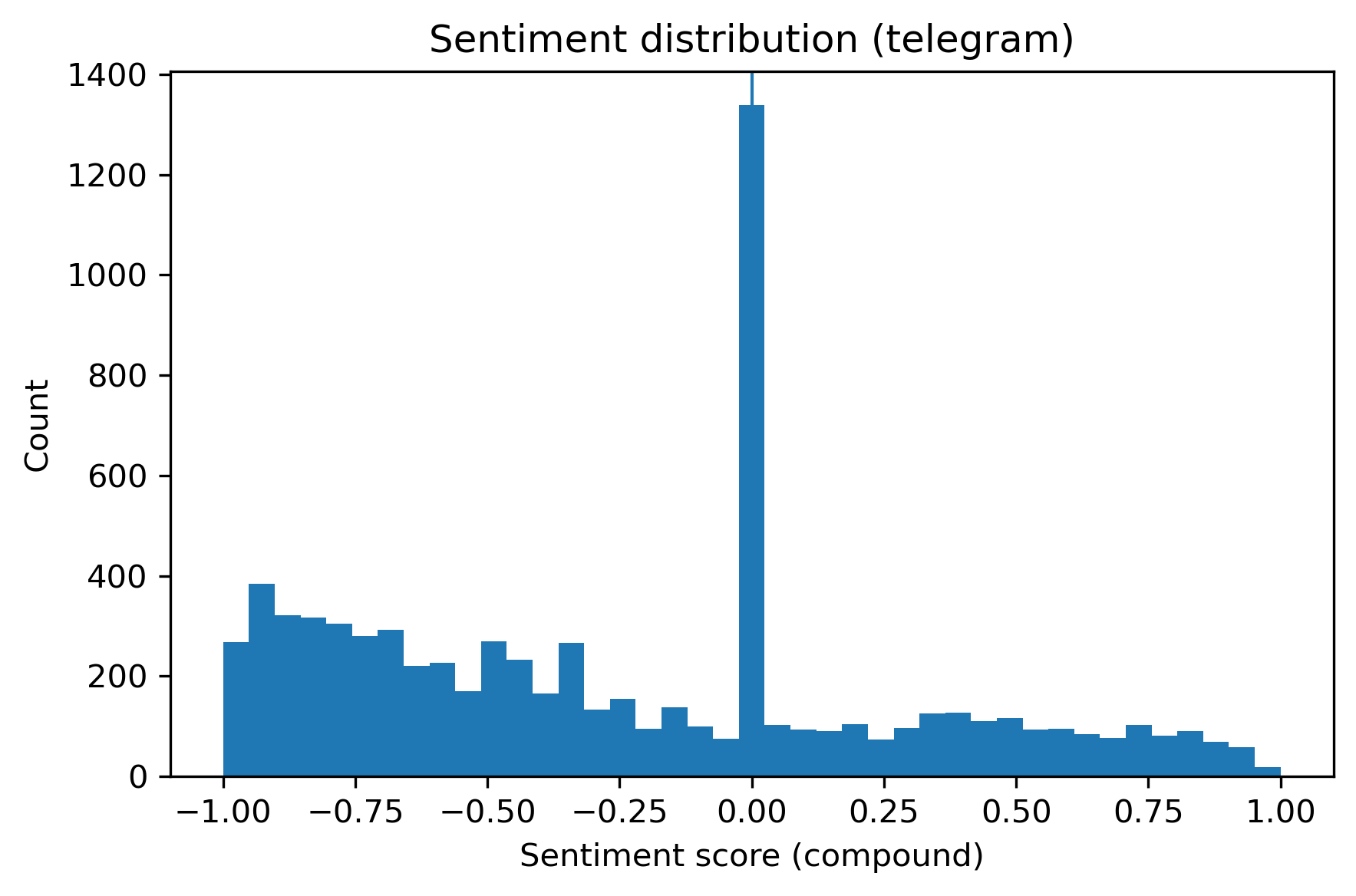}
        \caption{
        Distribution of sentiment scores for Telegram messages.
        The distribution is strongly skewed toward negative values, indicating persistent negative narrative tone and escalation-oriented discourse.
        }
        \label{fig:sentiment_distribution_telegram}
        \end{figure}

Figure~\ref{fig:sentiment_smoothed} shows the temporal evolution of sentiment across Telegram and Reddit using smoothed daily averages (14-day rolling mean), computed only for days with at least 10 messages.
Reddit exhibits predominantly neutral to moderately positive sentiment, typically between 0.20 and 0.40. Toward the end of the observation window, Reddit sentiment declines toward neutral, suggesting increasing uncertainty or concern in discussions.
In contrast, Telegram exhibits persistently negative sentiment throughout the entire period, typically between -0.20 and -0.35. This sustained negative polarity suggests that Telegram narratives are structurally more adversarial or escalation-oriented, consistent with its broadcast-style news framing.
Overall, the divergence indicates substantial platform differences in narrative tone: Telegram appears more aligned with threat- or escalation-centric framing, while Reddit remains closer to neutral/mixed sentiment.
Figures~\ref{fig:sentiment_distribution_reddit} and \ref{fig:sentiment_distribution_telegram} further confirm these differences through sentiment score distributions. Telegram shows a pronounced negative skew, whereas Reddit exhibits a more balanced distribution with a larger positive mass.

        Figures~\ref{fig:sentiment_distribution_reddit} and \ref{fig:sentiment_distribution_telegram} illustrate the structural differences in sentiment polarity between platforms.    

\subsection{Platform-Specific Topic and Event Signals}
To reduce cross-platform mixing and to validate whether temporal spikes reflect broad attention shifts rather than single-source activity, we report topic modeling and lexicon-based keyword activity separately for Telegram and Reddit.
\subsubsection{Telegram}

\begin{figure}[H]
\centering
\includegraphics[width=0.95\linewidth]{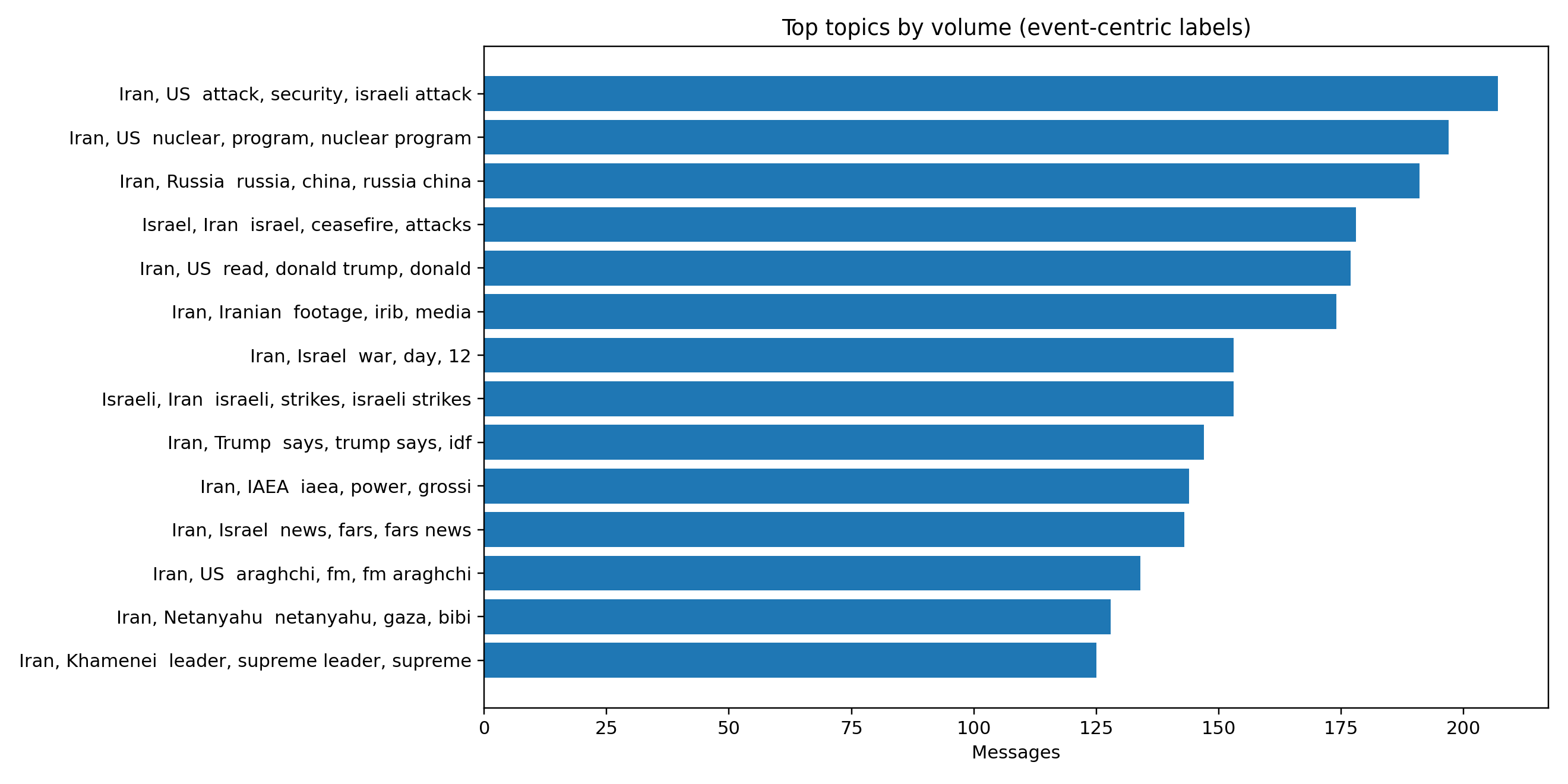}
\caption{Message volume by topic (Telegram only, NMF). Topics are labeled using representative keywords.}
\label{fig:tg_topic_volumes}
\end{figure}

\begin{figure}[H]
\centering
\includegraphics[width=0.95\linewidth]{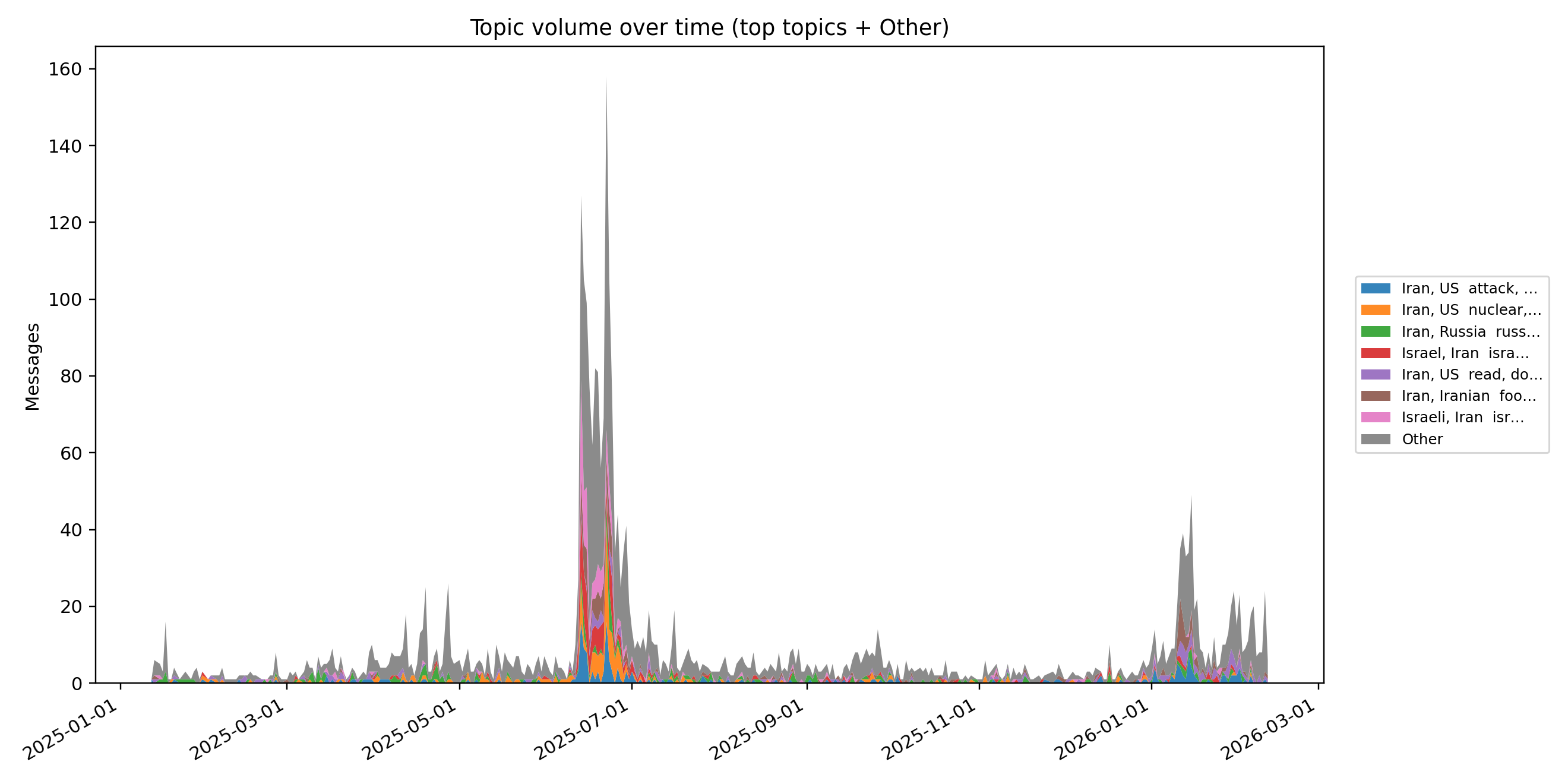}
\caption{Topic volume over time (Telegram only) for the top topics plus an \emph{Other} category.}
\label{fig:tg_topic_timeseries}
\end{figure}

Figure~\ref{fig:tg_topic_volumes} shows Telegram topic prevalence, dominated by broadcast-style reporting and recurring geopolitical frames. Temporal dynamics for the top topics (with remaining topics aggregated as \emph{Other}) are shown in Figure~\ref{fig:tg_topic_timeseries}.
Topic similarity is visualized using cosine-similarity neighborhood graphs (see Figure~\ref{fig:topic_network_tg} for Telegram and Figure~\ref{fig:topic_network_all} for the combined corpus). Dense connectivity indicates overlapping vocabulary across topics, while isolated nodes indicate more distinct themes.

\begin{figure}[H]
\centering
\includegraphics[width=0.95\linewidth]{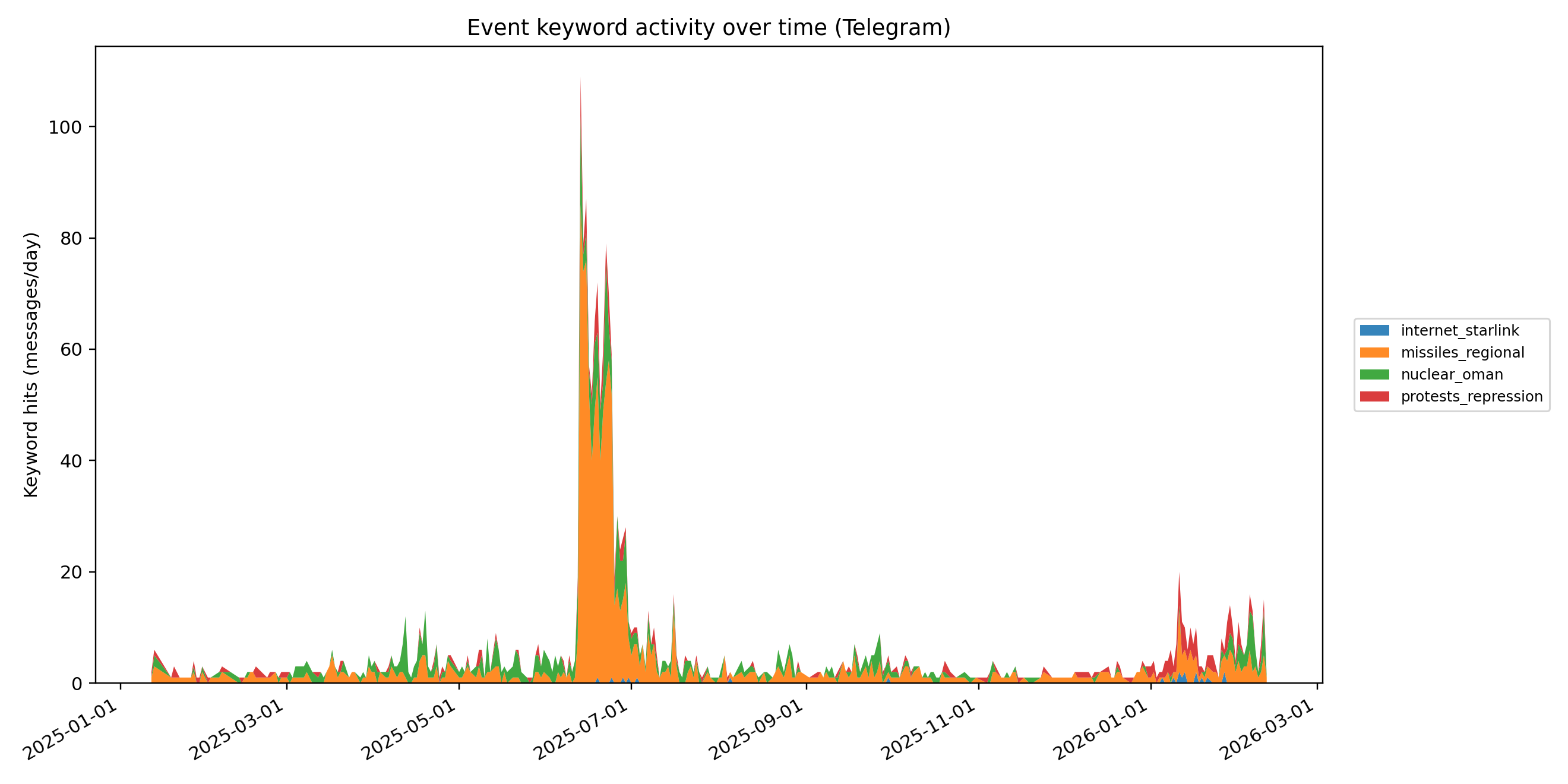}
\caption{Lexicon-based keyword activity over time (Telegram only, messages/day). Counts represent messages matching each keyword bundle, not unique events.}
\label{fig:tg_keyword_hits}
\end{figure}

Finally, Figure~\ref{fig:tg_keyword_hits} reports lexicon-based keyword activity over time; these counts represent messages matching keyword bundles and should be interpreted as an attention proxy rather than unique events.

\subsubsection{Reddit}
\begin{figure}[H]
\centering
\includegraphics[width=0.95\linewidth]{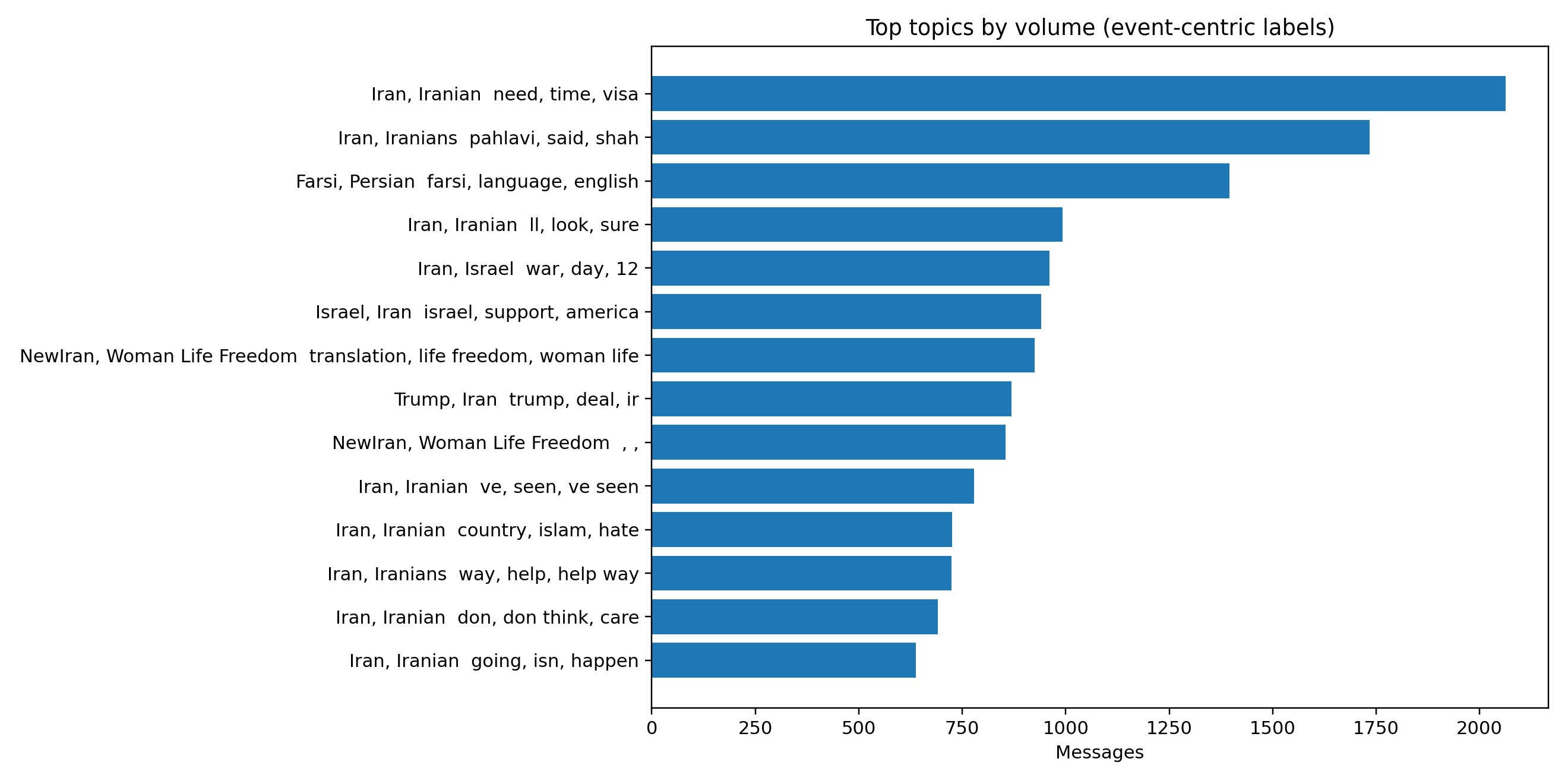}
\caption{Message volume by topic (Reddit only, NMF). Topics are labeled using representative keywords.}
\label{fig:rd_topic_volumes}
\end{figure}

\begin{figure}[H]
\centering
\includegraphics[width=0.95\linewidth]{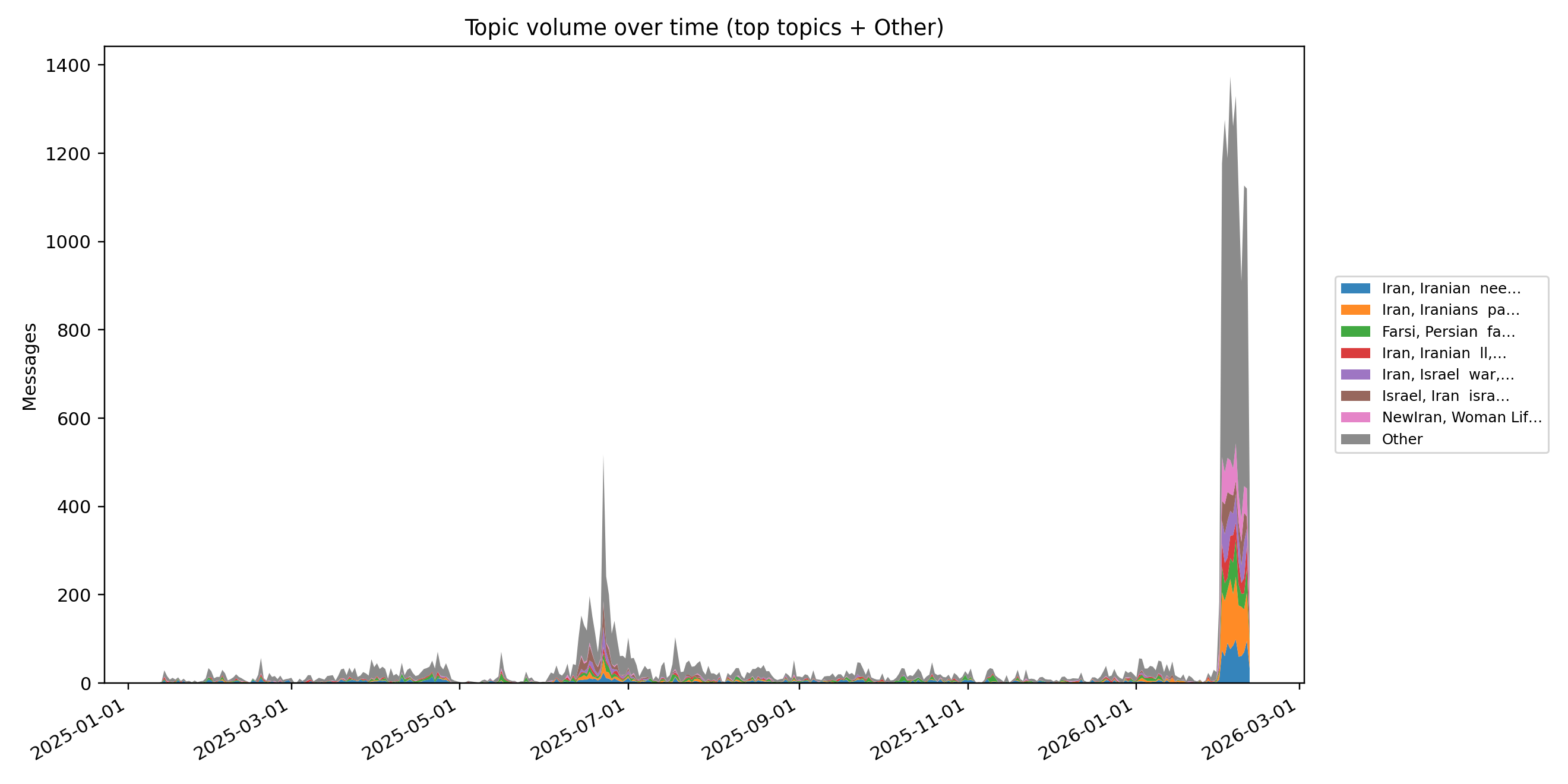}
\caption{Topic volume over time (Reddit only) for the top topics plus an \emph{Other} category.}
\label{fig:rd_topic_timeseries}
\end{figure}

\begin{figure}[H]
\centering
\includegraphics[width=0.95\linewidth]{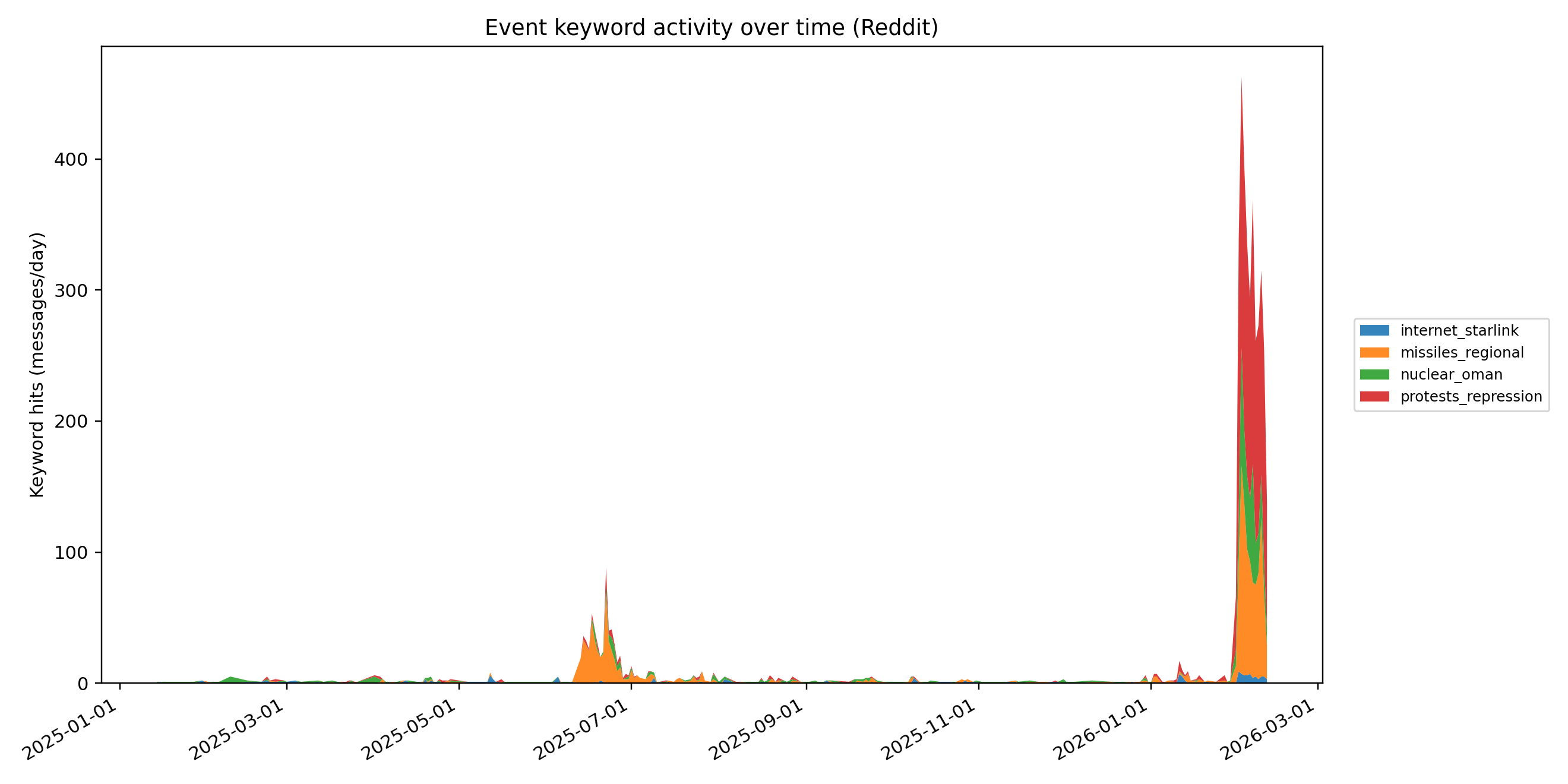}
\caption{Lexicon-based keyword activity over time (Reddit only, messages/day). Counts represent messages matching each keyword bundle, not unique events.}
\label{fig:rd_keyword_hits}
\end{figure}

Compared to Telegram, Reddit exhibits more conversational and community-driven framing, which affects topic structure and overlap. 
Figure~\ref{fig:rd_topic_volumes} shows topic prevalence, while temporal dynamics are shown in Figure~\ref{fig:rd_topic_timeseries}. 
Topic similarity is visualized using cosine-similarity neighborhood graphs (see Figure~\ref{fig:topic_network_rd} for Reddit and Figure~\ref{fig:topic_network_all} for the combined corpus). Dense connectivity indicates overlapping vocabulary across topics, while isolated nodes indicate more distinct themes. Figure~\ref{fig:rd_keyword_hits} reports lexicon-based keyword activity over time; as with Telegram, these are message-level matches and should be interpreted as an attention proxy rather than unique events.

Overlap can be stronger on Reddit due to shared conversational markers and recurring community jargon, which encourages cross-topic blending.

\subsection{External Event Validation}

To validate whether observed discourse spikes correspond to real-world events, we compared topic and keyword activity timelines with an external dataset of daily protest statistics. This comparison enables assessment of whether online discourse responds to offline political developments \url{https://www.kaggle.com/datasets/justin2028/daily-statistics-of-the-2022-iran-protests}.

\subsubsection{Entity Co-occurrence Network}
\label{subsec:entity_network}

To capture the structural context of discourse and identify which geopolitical actors co-appear in narrative framing, we construct an entity co-occurrence network from named entities (PERSON/ORG/GPE). Nodes represent entities and edges indicate co-occurrence within the same message or post. Node size is proportional to entity frequency and edge thickness reflects co-occurrence strength.

\begin{figure}[H]
\centering
\includegraphics[width=0.90\linewidth]{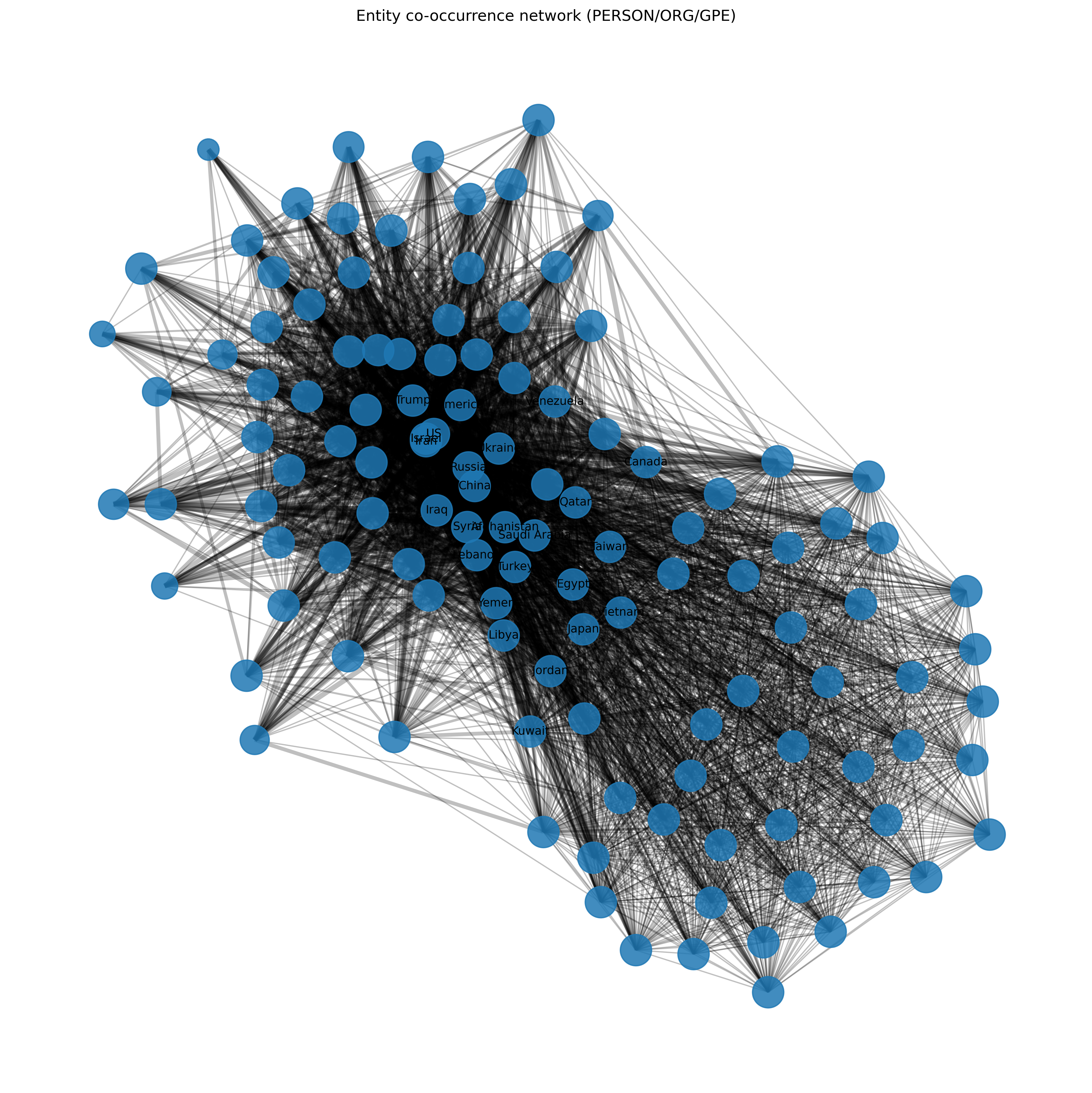}
\caption{Entity co-occurrence network extracted from Telegram and Reddit messages using named entity recognition (PERSON/ORG/GPE). Nodes represent entities and edges indicate co-occurrence within the same message/post. Node size is proportional to entity frequency and edge thickness reflects co-occurrence strength. The dense core highlights dominant actors and locations frequently mentioned together, providing structural context for the validation of escalation-related narratives.}
\label{fig:entity_network_all}
\end{figure}

Figure~\ref{fig:entity_network_all} reveals a dense core of frequently co-mentioned actors and locations, indicating that Iran-related narratives are embedded in a broader geopolitical context involving regional and international stakeholders. This supports interpretation of escalation signals as multi-actor narratives rather than isolated keyword bursts and complements the temporal validation analysis.

\begin{figure}[ht]
    \centering
    \includegraphics[width=0.9\linewidth]{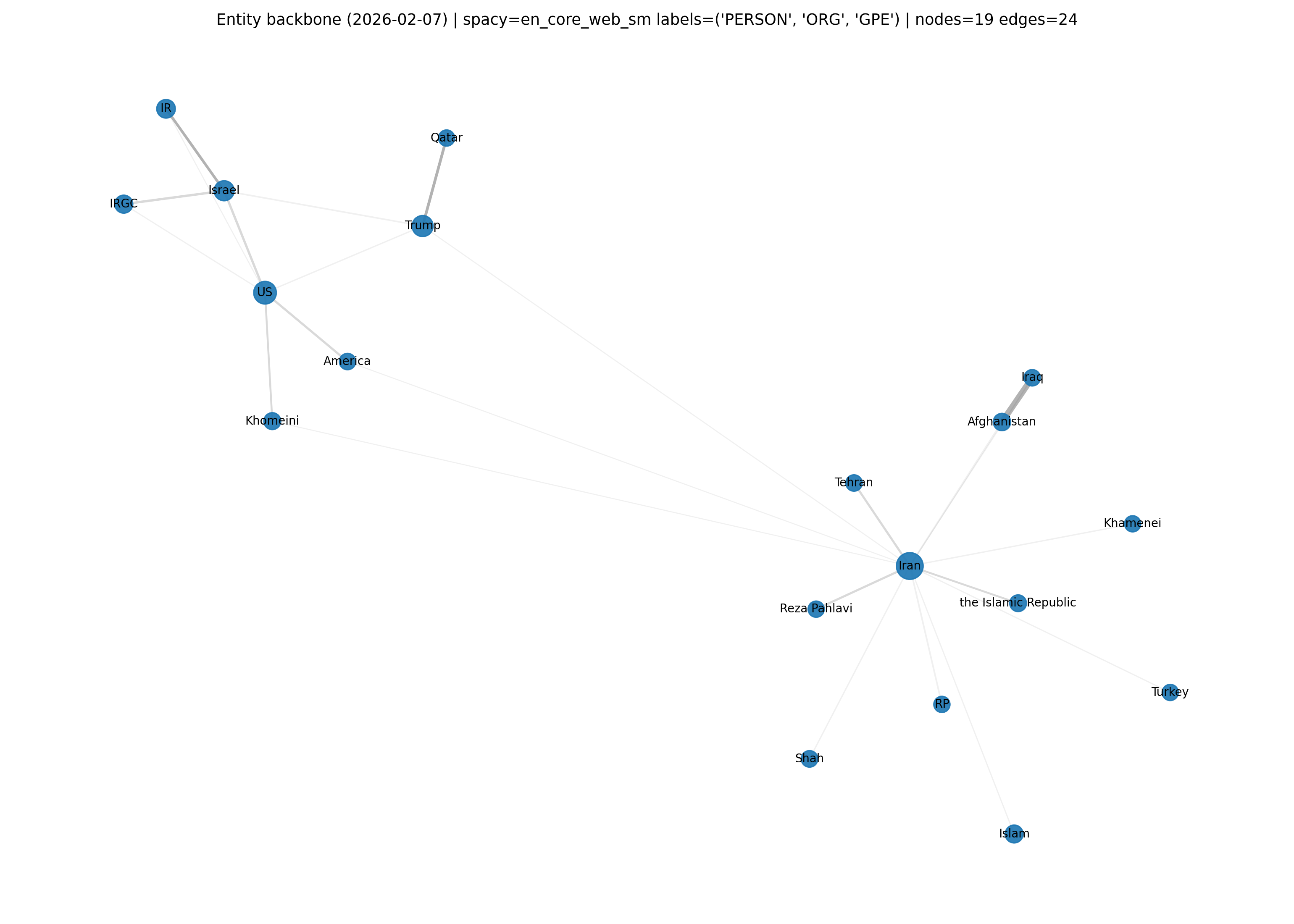}
    \caption{Entity co-occurrence backbone for 7 February 2026. Nodes represent named entities (PERSON, ORG, GPE) extracted using spaCy. Edge weights are PMI-based and pruned using a top-$k$ backbone filter to retain the strongest co-occurrence relationships. Node size reflects entity frequency within the filtered window. The network contains 19 nodes and 24 edges.}
    \label{fig:entity_backbone_20260207}
    \end{figure}

Figure~\ref{fig:entity_backbone_20260207} presents the entity co-occurrence backbone for 7 February 2026, constructed using PMI-weighted co-occurrence and backbone pruning. The resulting network (19 nodes, 24 edges) reveals a clearly centralized structure anchored around \textit{Iran}, which functions as the dominant hub connecting multiple geopolitical actors and institutional references. Surrounding Iran are entities such as \textit{Tehran}, \textit{the Islamic Republic}, \textit{Khamenei}, \textit{Reza Pahlavi}, and \textit{Shah}, indicating simultaneous discussion of contemporary regime politics and historical-monarchical references. 

A secondary cluster connects \textit{US}, \textit{Israel}, \textit{IRGC}, and \textit{IR}, reflecting security and regional conflict framing. The linkage of \textit{Trump} and \textit{Qatar} to this cluster suggests U.S.-centric geopolitical positioning within the discourse. Additional peripheral links to \textit{Iraq}, \textit{Afghanistan}, and \textit{Turkey} highlight the broader regional embedding of Iran-related narratives.

Overall, the backbone structure demonstrates that discourse on this date was organized around a small set of tightly interconnected geopolitical actors, with Iran serving as the central narrative anchor. By filtering weaker co-mentions, the backbone visualization reveals structurally coherent narrative clusters centered on regime legitimacy, regional security dynamics, and U.S.–Israel–Iran tensions.

\section{Protest Event Validation Using Twitter Data}
\label{sec:twitter_event_validation}

To evaluate whether Twitter discourse reflects real-world protest dynamics, we performed a correlation analysis between daily protest counts and daily keyword hit frequencies derived from Twitter data. This analysis examines both same-day relationships and lagged relationships to determine whether Twitter activity precedes, coincides with, or follows protest events.

\begin{figure}[H]
\centering
\includegraphics[width=0.9\linewidth]{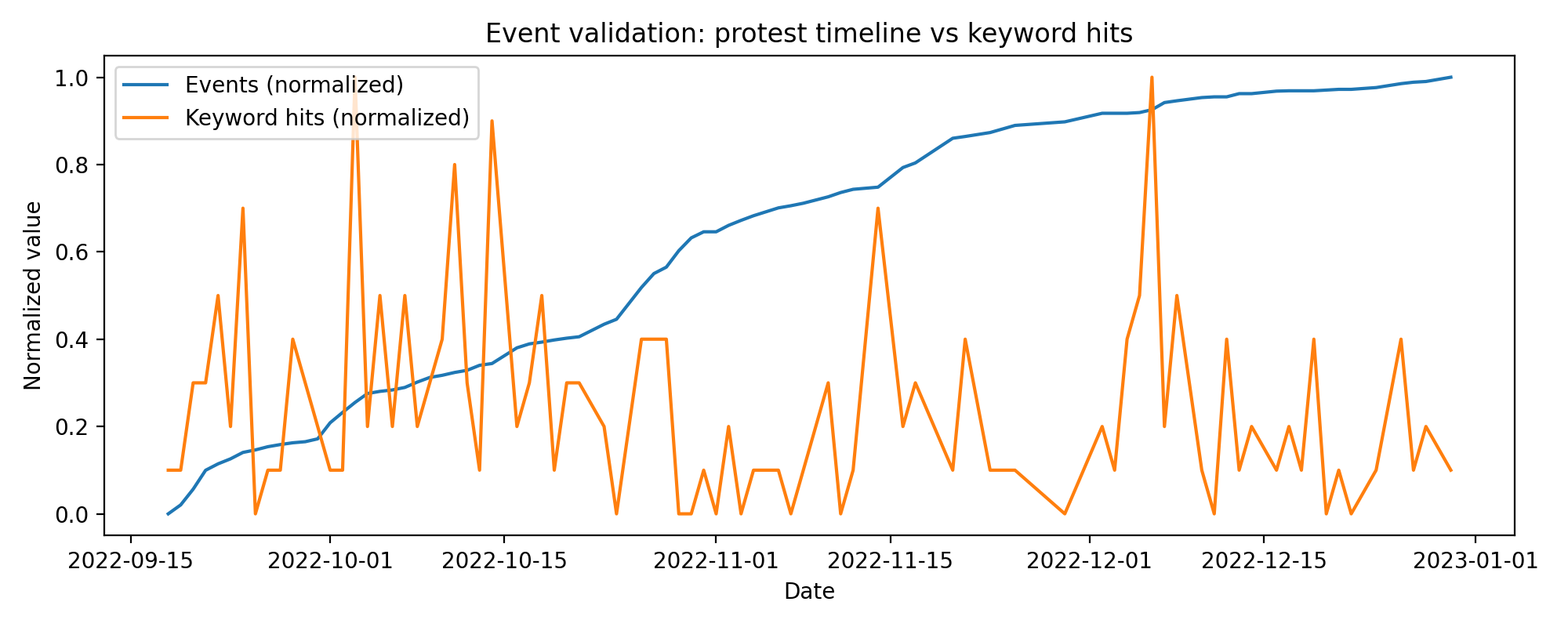}
\caption{Normalized timeline comparison between protest events and Twitter keyword hits. While protest counts show a steady increase, keyword hits exhibit irregular spikes that do not consistently align with protest activity.}
\label{fig:event_timeline}
\end{figure}

Figure~\ref{fig:event_timeline} presents the normalized timelines of protest events and keyword hits between September 17, 2022 and December 31, 2022. The protest event series exhibits a generally increasing trend during the observation period, while keyword hits show intermittent spikes rather than a continuous upward trend. These spikes appear temporally uneven and do not consistently align with increases in protest event counts.

\begin{figure}[H]
\centering
\includegraphics[width=0.8\linewidth]{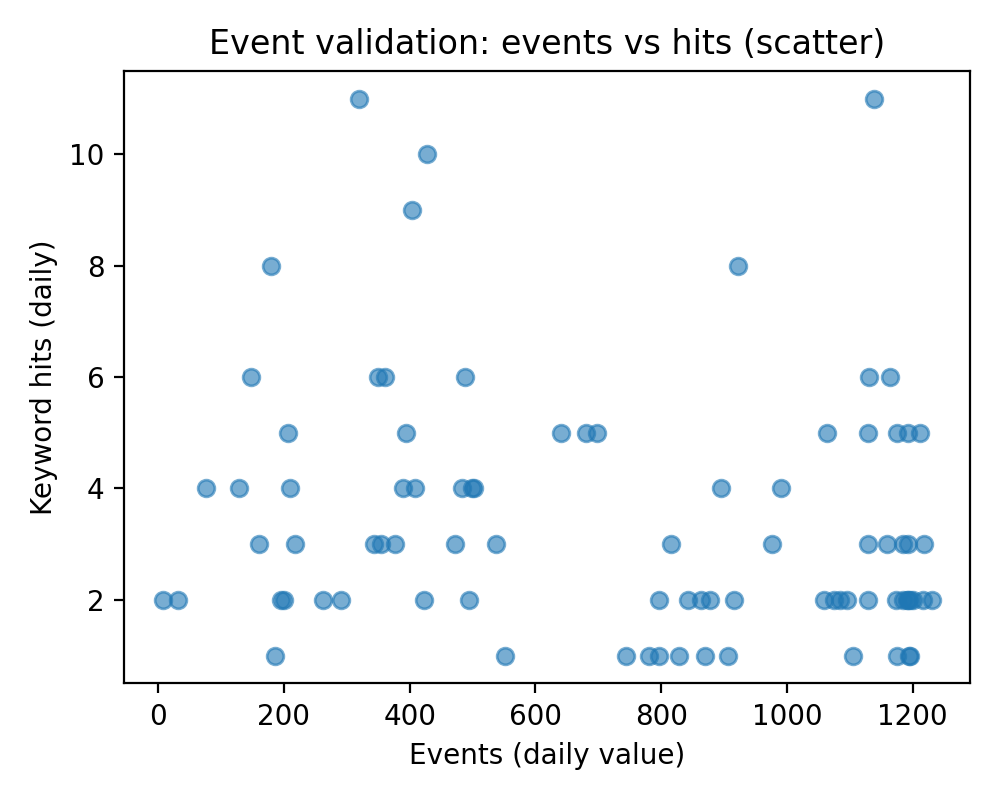}
\caption{Scatter plot of daily protest counts versus Twitter keyword hits. The lack of a clear trend confirms the weak correlation between the two variables.}
\label{fig:event_scatter}
\end{figure}

Figure~\ref{fig:event_scatter} shows the scatter plot of daily protest counts versus daily keyword hit frequency. The distribution of points does not reveal a clear linear or monotonic relationship. This observation is supported by the correlation analysis, which yielded weak negative same-day correlations (Pearson $r=-0.19$, Spearman $\rho=-0.21$). These results indicate that Twitter keyword frequency does not directly mirror daily protest intensity.

\begin{figure}[H]
\centering
\includegraphics[width=0.9\linewidth]{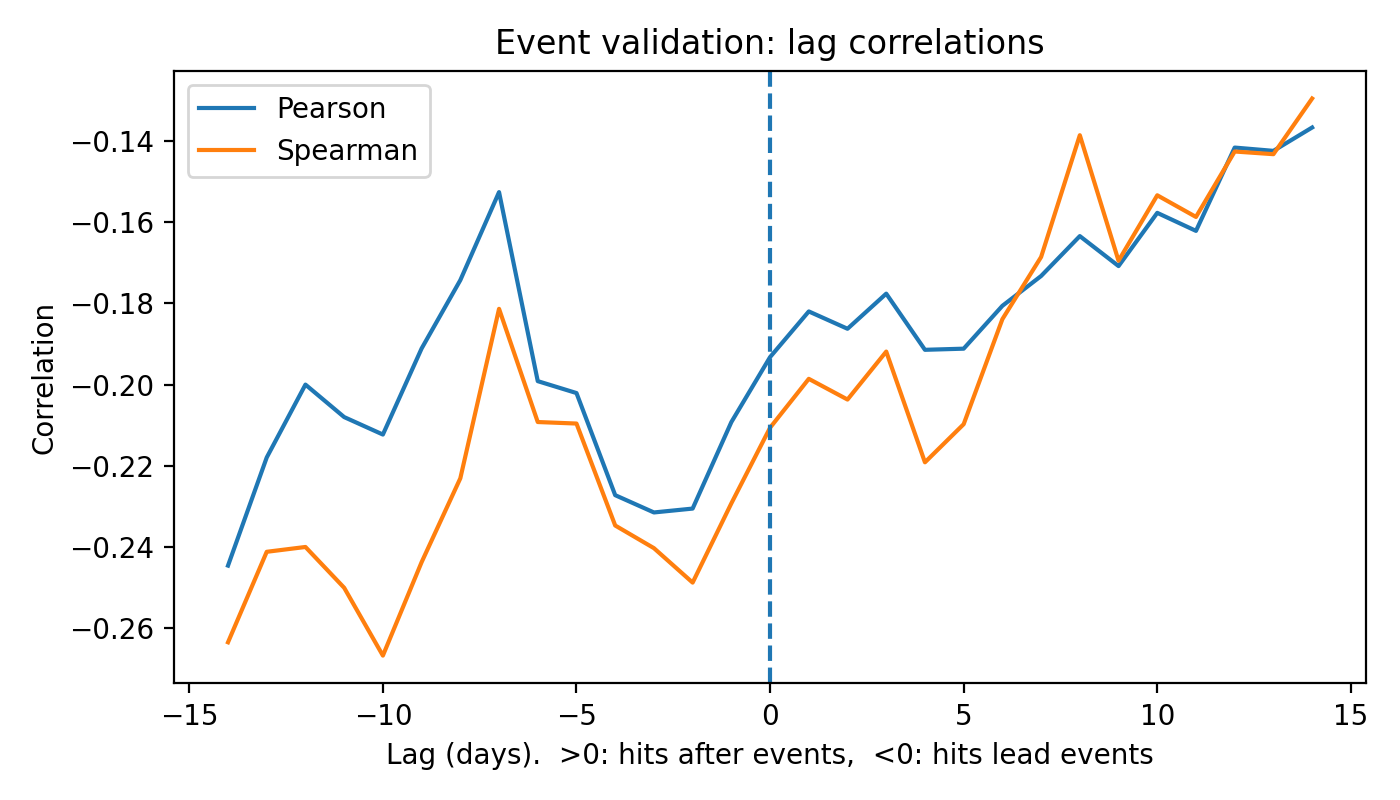}
\caption{Lagged correlation between protest events and keyword hits. Negative lag values correspond to keyword activity preceding protest events. The strongest correlations occur at lags of approximately $-10$ to $-14$ days.}
\label{fig:event_lags}
\end{figure}

To further investigate temporal dynamics, we computed lagged correlations between the two time series, shown in Figure~\ref{fig:event_lags}. Negative lag values indicate that keyword hits precede protest events, while positive lag values indicate that keyword hits follow protest events. The strongest correlations were observed at negative lags of approximately $-10$ to $-14$ days (Spearman $\rho=-0.27$, Pearson $r=-0.24$). This suggests that increases in Twitter keyword activity may occur prior to increases in protest events, potentially reflecting early mobilization, anticipation, or pre-event discourse.

Overall, the results suggest that Twitter keyword activity does not strongly track protest counts on a day-to-day basis. Instead, Twitter discourse may function as an early signal or indicator of protest mobilization rather than a direct reflection of protest intensity.

%\section{Real-Time Narrative Escalation and Correlation with Geopolitical Events}
\section{Detection of Narrative Escalation and Correlation with Geopolitical Events}
\label{sec:realtime_escalation}

To evaluate whether the proposed analytical framework can detect emerging geopolitical developments in real time, we collected an additional dataset from Reddit and Telegram on February 25, 2026. This real-time acquisition was motivated by increased Iran-related activity observed across major news-oriented subreddits such as \texttt{r/news} and \texttt{r/worldnews}, as well as Telegram channels associated with international news reporting.
This real-time subset was collected separately from the historical dataset described in Section~\ref{sec:dataset} and is used exclusively to evaluate escalation detection capability.

The real-time dataset comprises 4,955 Telegram messages and 49,014 Reddit items, including 947 posts and 48,067 comments. Telegram messages span January 14, 2018 to February 25, 2026, while Reddit items span June 22, 2021 to February 25, 2026. These data were collected using automated crawlers targeting Iran-related news and discussion communities. The dataset captures both baseline discourse and recent escalation-related activity, enabling assessment of emerging geopolitical narratives (Table~\ref{tab:realtime_dataset}).

\begin{table}[h]
\centering
\begin{tabular}{lrrrr}
\hline
Platform & Messages/Items & Posts & Comments & Date Range \\
\hline
Telegram & 4,955 & -- & -- & 2018-01-14 to 2026-02-25 \\
Reddit   & 49,014 & 947 & 48,067 & 2021-06-22 to 2026-02-25 \\
\hline
Total    & 53,969 & 947 & 48,067 & 2018-01-14 to 2026-02-25 \\
\hline
\end{tabular}
\caption{Real-time dataset used to evaluate escalation detection capability.}
\label{tab:realtime_dataset}
\end{table}

Although collected in February 2026, the Reddit and Telegram crawlers retrieve historical items when available through platform search interfaces and channel history. As a result, the real-time dataset includes both recent messages reflecting emerging geopolitical developments and earlier messages providing baseline discourse coverage. This combined temporal coverage enables reliable detection of escalation signals relative to historical activity levels.
\subsection{Escalation Signals in Message Volume}

\begin{figure}[H]
\centering
\includegraphics[width=\linewidth]{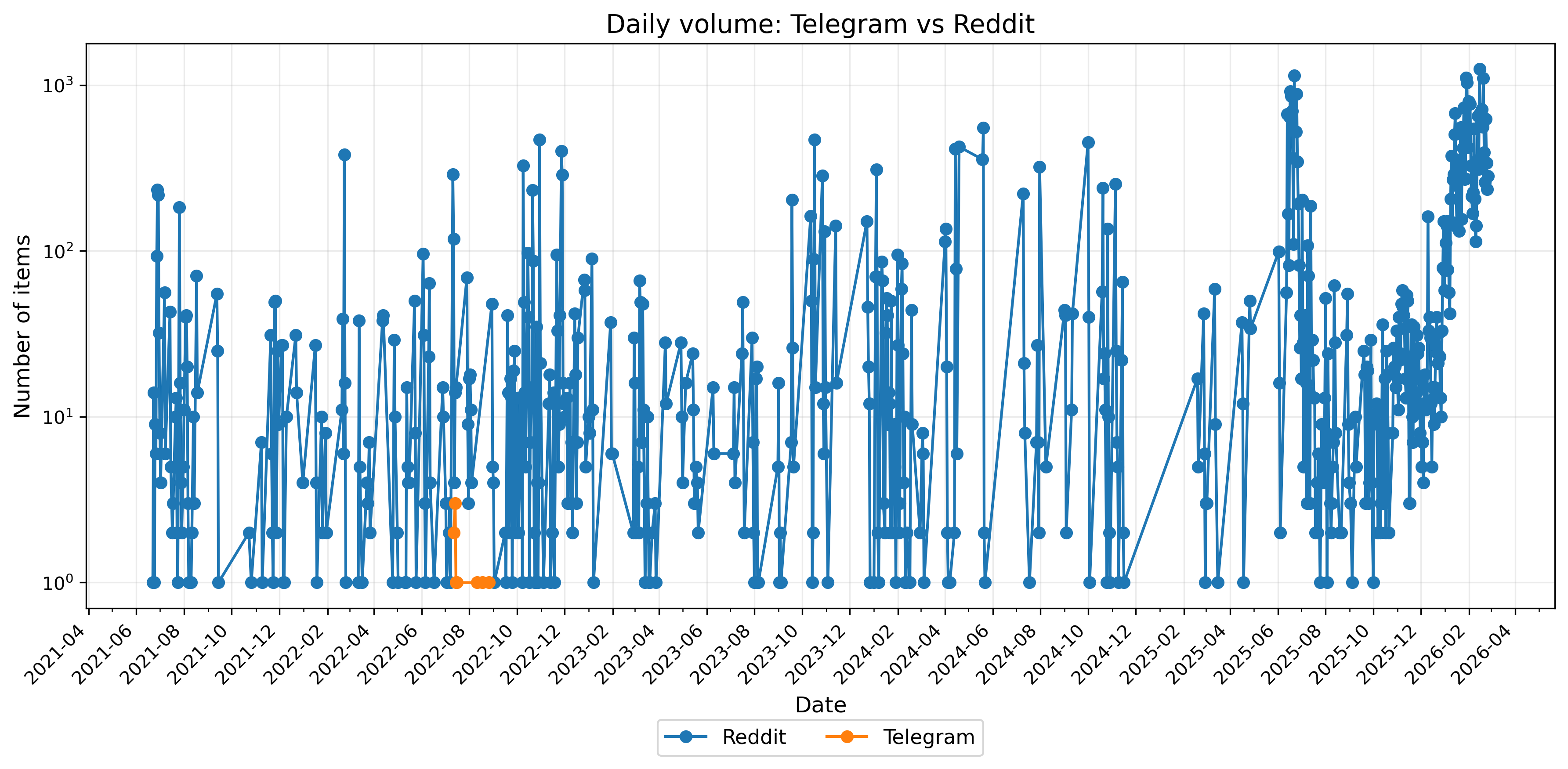}
\caption{Daily Iran-related message volume across Reddit and Telegram. Reddit exhibits a substantial increase in activity beginning in late 2025 and peaking in early 2026, indicating intensified attention and narrative amplification. Telegram activity remains lower but follows similar temporal patterns.}
\label{fig:realtime_volume}
\end{figure}

We first examine daily message volume as a coarse indicator of attention dynamics. Figure~\ref{fig:realtime_volume} shows a substantial increase in Reddit activity beginning in late 2025 and intensifying into early 2026. This increase reflects heightened public attention and active discussion related to Iran. Telegram activity remains lower in absolute volume due to the smaller number of monitored channels, but exhibits temporal alignment with escalation periods observed on Reddit. This cross-platform synchronization suggests that both broadcast media and discussion communities respond to emerging geopolitical developments.

\subsection{Escalation Signals in Narrative Content}

To identify escalation-specific discourse, we analyzed keyword bundles representing military, nuclear, diplomatic, and escalation-related terminology.

\begin{figure}[H]
\centering
\includegraphics[width=\linewidth]{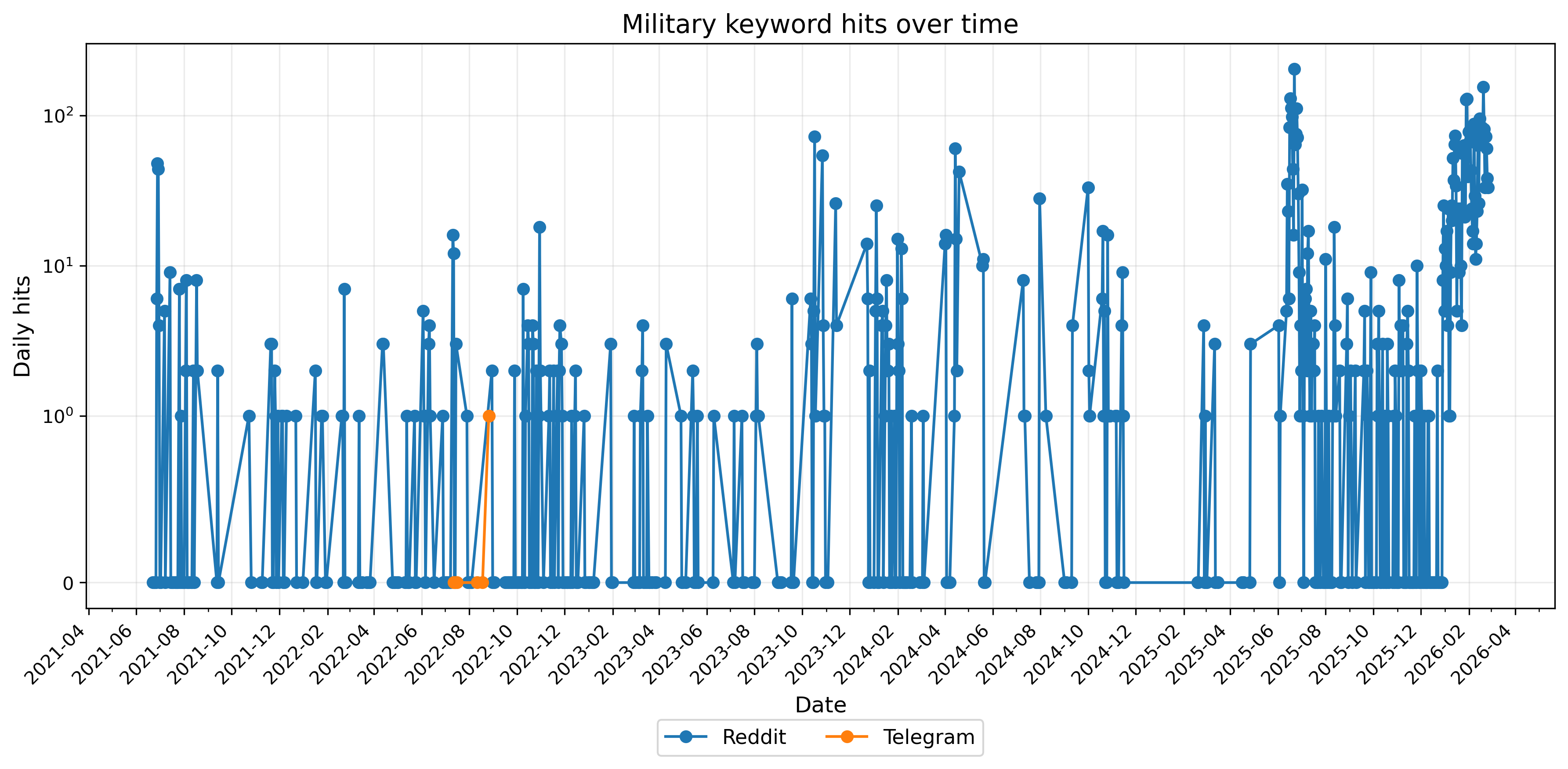}
\caption{Military-related keyword activity over time. Peaks correspond to increased discussion of military deployments, strategic positioning, and escalation-related developments.}
\label{fig:realtime_military}
\end{figure}

\begin{figure}[H]
\centering
\includegraphics[width=\linewidth]{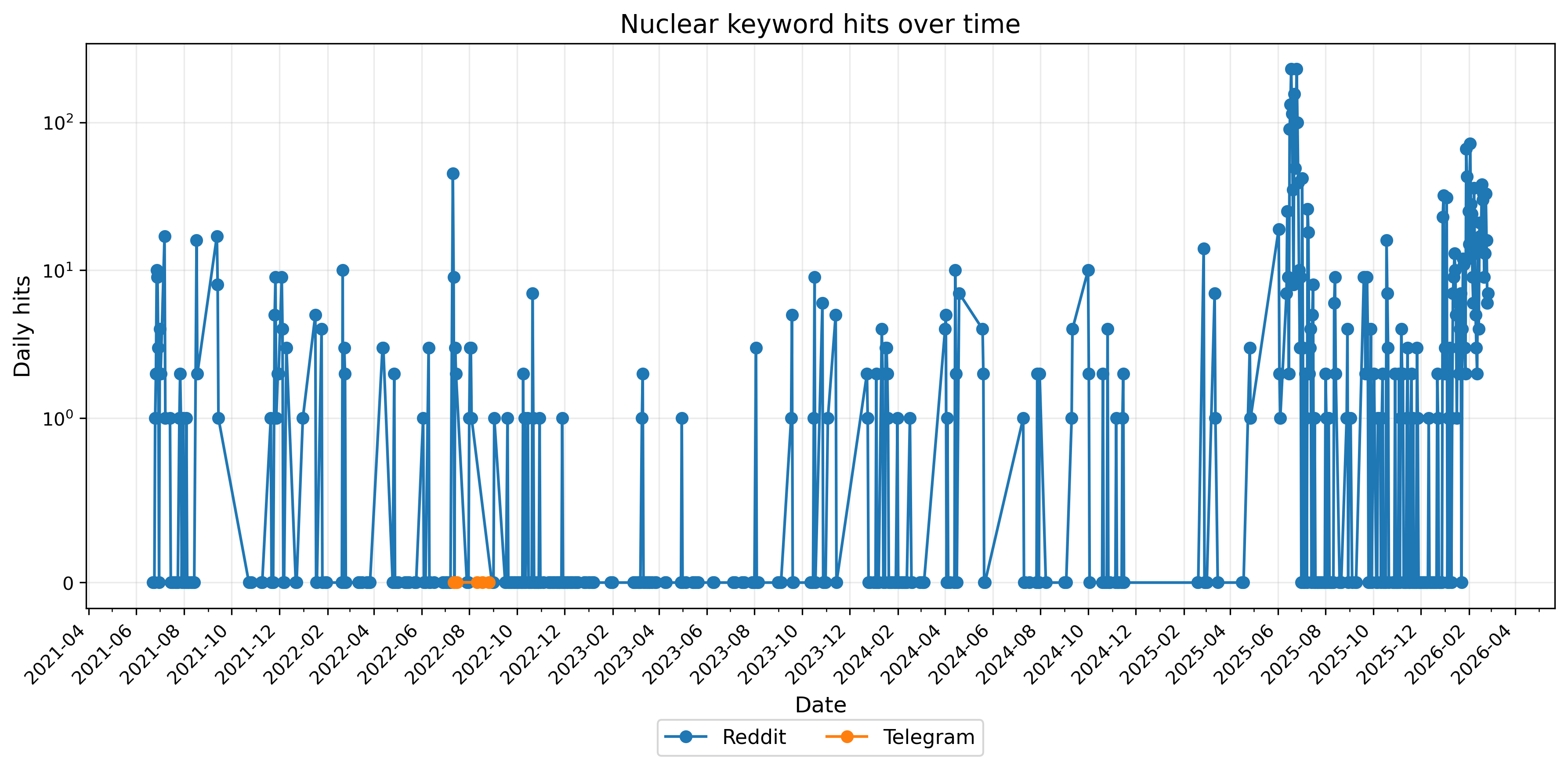}
\caption{Nuclear-related keyword activity over time, reflecting increased attention to nuclear policy, enrichment, and international regulatory concerns.}
\label{fig:realtime_nuclear}
\end{figure}

\begin{figure}[H]
\centering
\includegraphics[width=\linewidth]{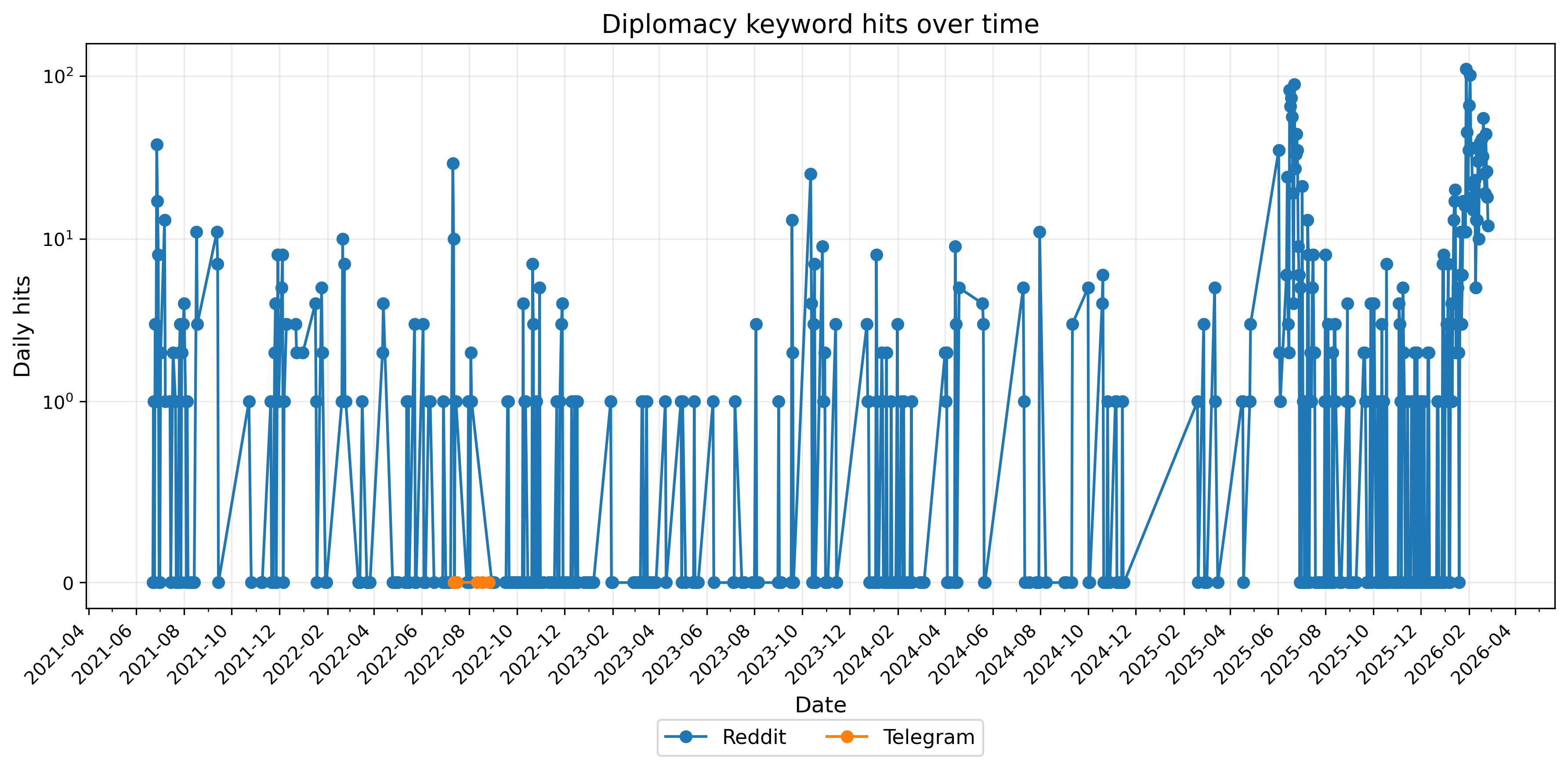}
\caption{Diplomatic keyword activity over time. Increased activity reflects heightened diplomatic engagement and international response during escalation periods.}
\label{fig:realtime_diplomacy}
\end{figure}

Military-related discourse (Figure~\ref{fig:realtime_military}) exhibits clear spikes during escalation periods, indicating increased discussion of military deployments, strategic positioning, and potential conflict scenarios. Nuclear-related discourse similarly increases (Figure~\ref{fig:realtime_nuclear}), reflecting renewed attention to nuclear policy, enrichment, and international regulatory concerns. Diplomatic discourse also intensifies (Figure~\ref{fig:realtime_diplomacy}), indicating active international engagement, negotiations, and geopolitical response.

\subsection{Composite Escalation Index}

To quantify escalation intensity, we construct a composite escalation index by aggregating normalized keyword bundle frequencies across military, nuclear, diplomatic, and escalation-specific terms. This index provides a unified measure of escalation-related narrative activity across platforms.

\begin{figure}[H]
\centering
\includegraphics[width=\linewidth]{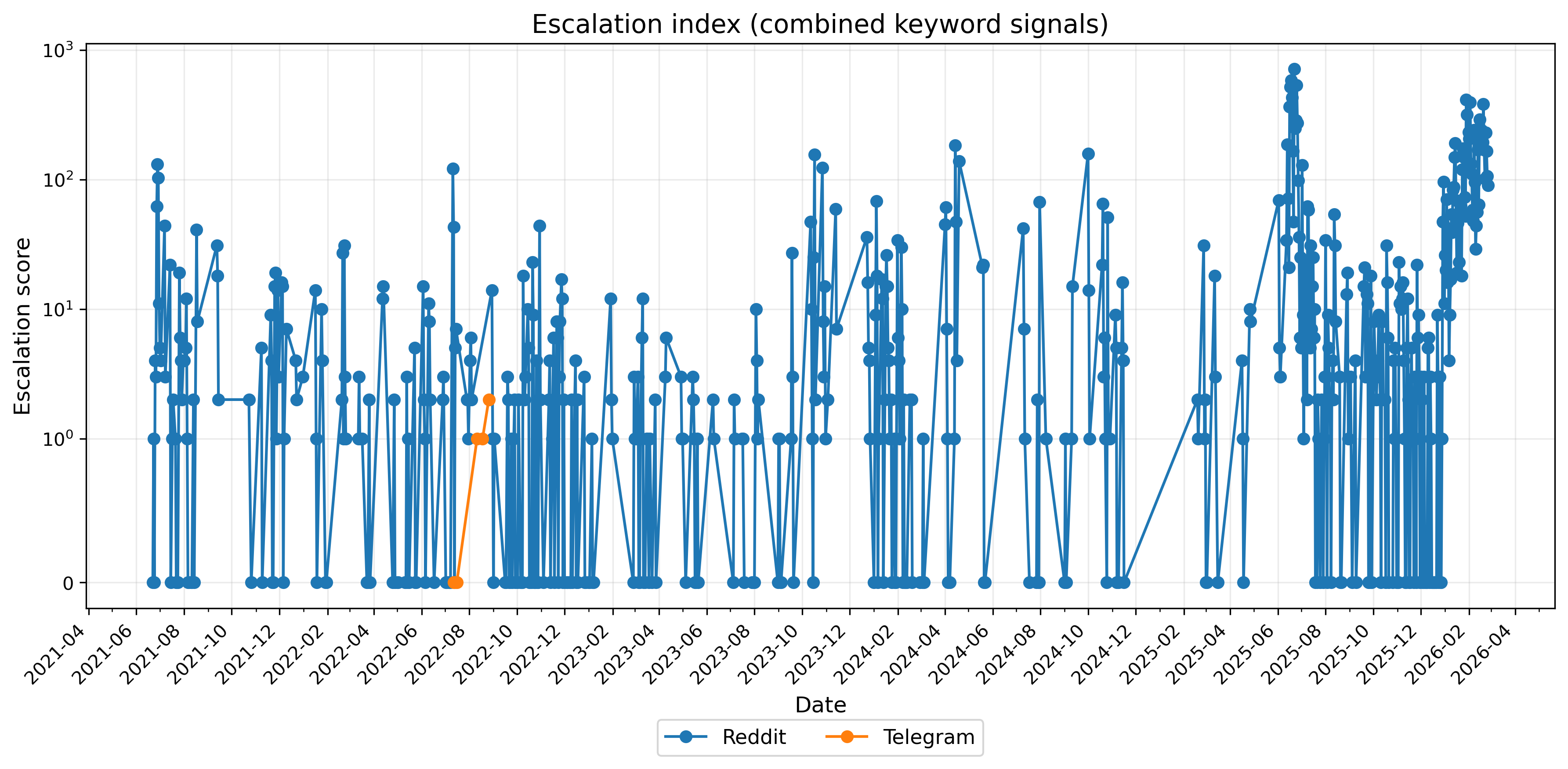}
\caption{Composite escalation index derived from combined keyword bundle frequencies. The index shows a substantial increase in early 2026, indicating elevated escalation-related narrative activity across platforms.}
\label{fig:realtime_escalation_index}
\end{figure}

As shown in Figure~\ref{fig:realtime_escalation_index}, the escalation index exhibits a pronounced increase beginning in late 2025 and peaking in early 2026. This increase reflects simultaneous amplification across multiple escalation-related narrative dimensions, indicating coordinated narrative escalation rather than isolated keyword fluctuations.

\subsection{Correlation with Documented Geopolitical Events}

To assess whether online escalation signals correspond to real-world geopolitical developments, we compared the escalation index with a curated timeline of documented geopolitical events involving Iran.

\begin{figure}[H]
\centering
\includegraphics[width=\linewidth]{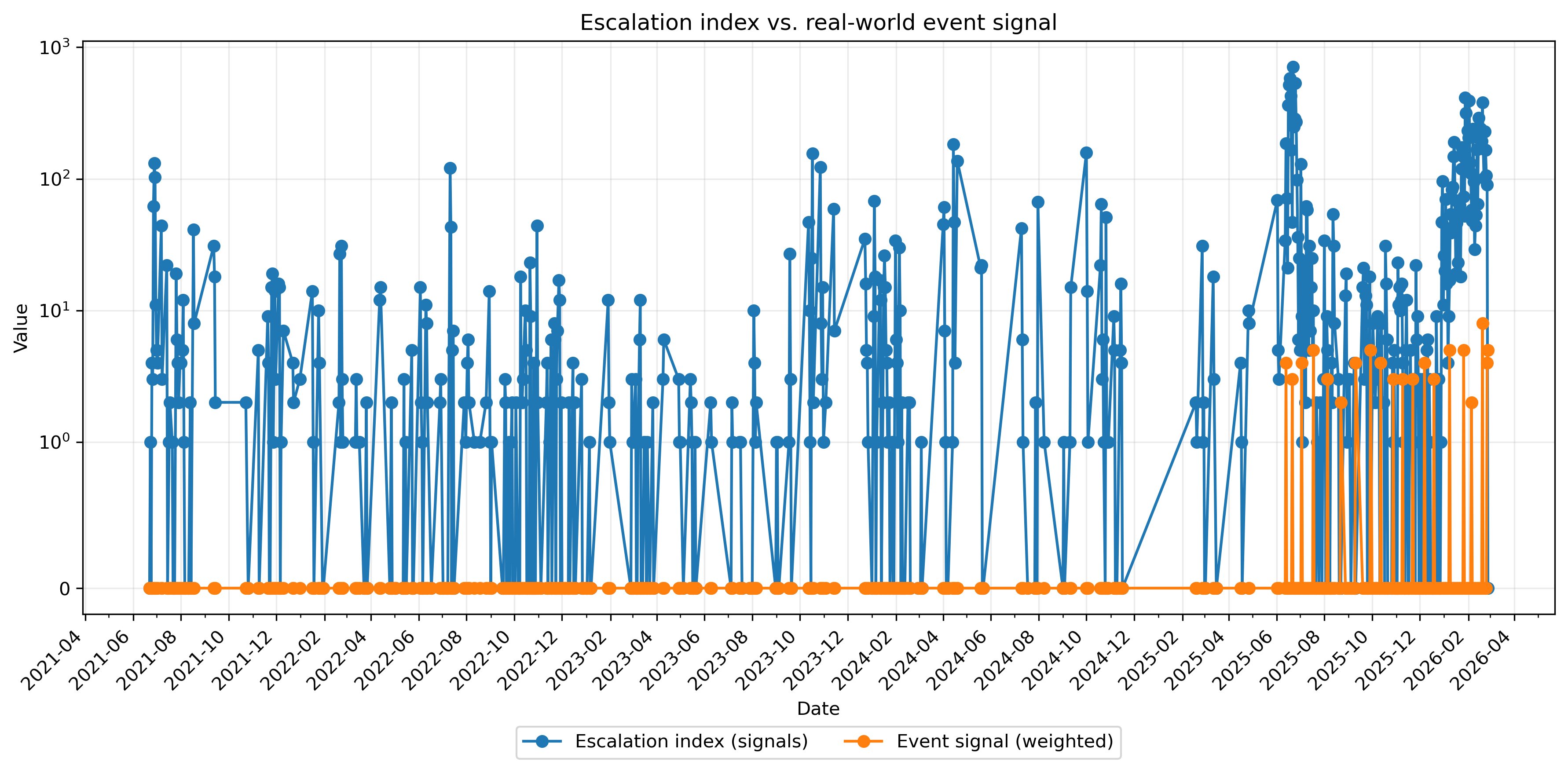}
\caption{Comparison between escalation index and documented geopolitical events. Peaks in escalation index align with real-world geopolitical developments, indicating correspondence between online discourse and geopolitical activity.}
\label{fig:events_vs_escalation}
\end{figure}

Figure~\ref{fig:events_vs_escalation} shows clear temporal alignment between escalation index peaks and documented geopolitical developments, including military positioning, diplomatic announcements, and nuclear-related developments. This alignment indicates that narrative escalation observed in online discourse reflects real-world geopolitical dynamics.

\begin{figure}[H]
\centering
\includegraphics[width=\linewidth]{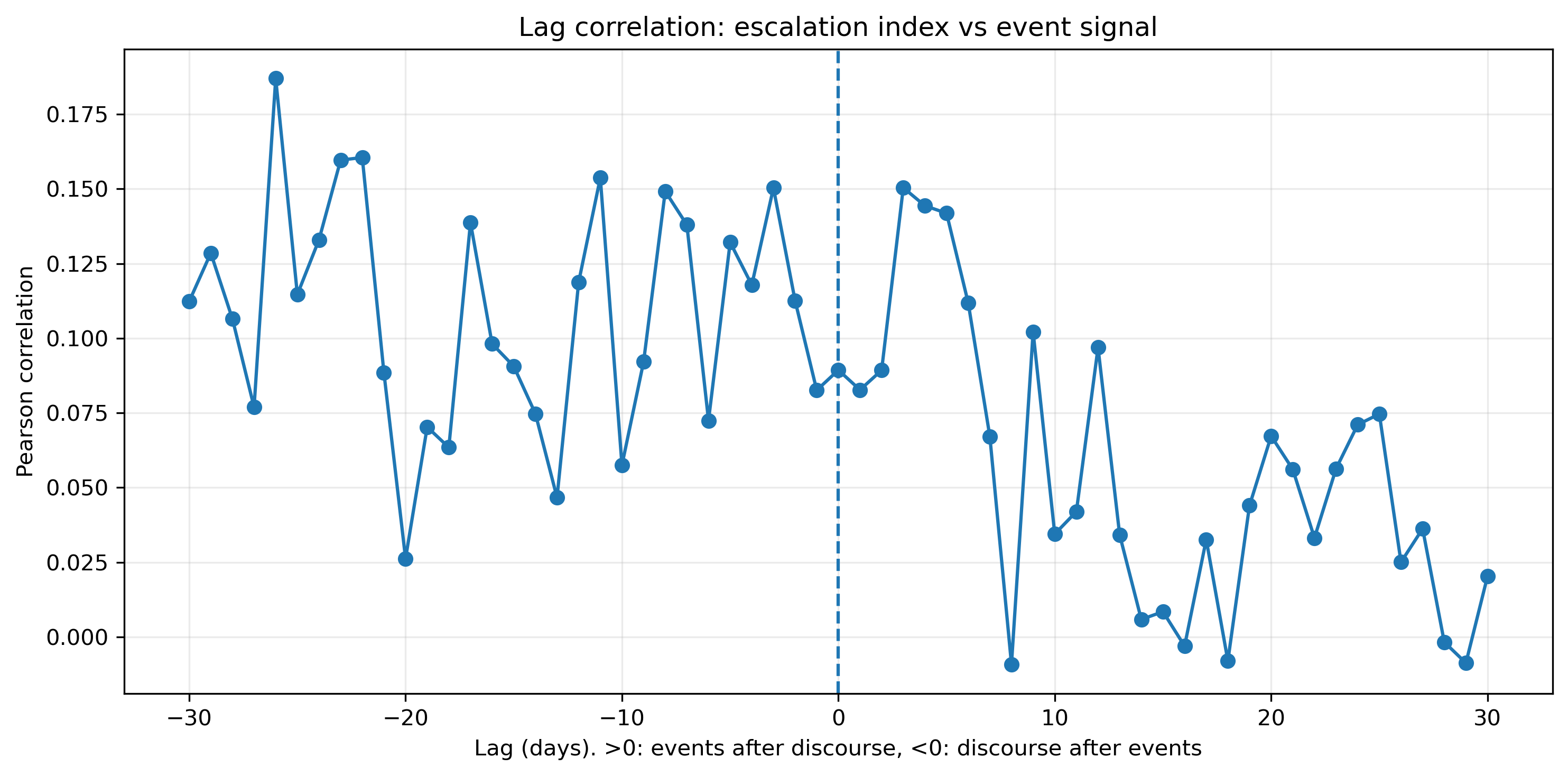}
\caption{Lagged correlation between escalation index and documented geopolitical events. Correlations across multiple temporal offsets demonstrate both reactive and anticipatory relationships between online discourse and real-world developments.}
\label{fig:lag_correlation}
\end{figure}

To quantify this relationship, we computed lagged Pearson correlation values between the escalation index and event signal across temporal offsets. Figure~\ref{fig:lag_correlation} shows meaningful correlations across multiple lags. Negative lag correlations indicate reactive narrative amplification following geopolitical events, while positive lag correlations indicate that escalation discourse may sometimes precede formally documented events, potentially reflecting early reporting, anticipation, or emerging information dissemination.

\subsection{Structural Context of Escalation Narratives}

\begin{figure}[H]
\centering
\includegraphics[width=\linewidth]{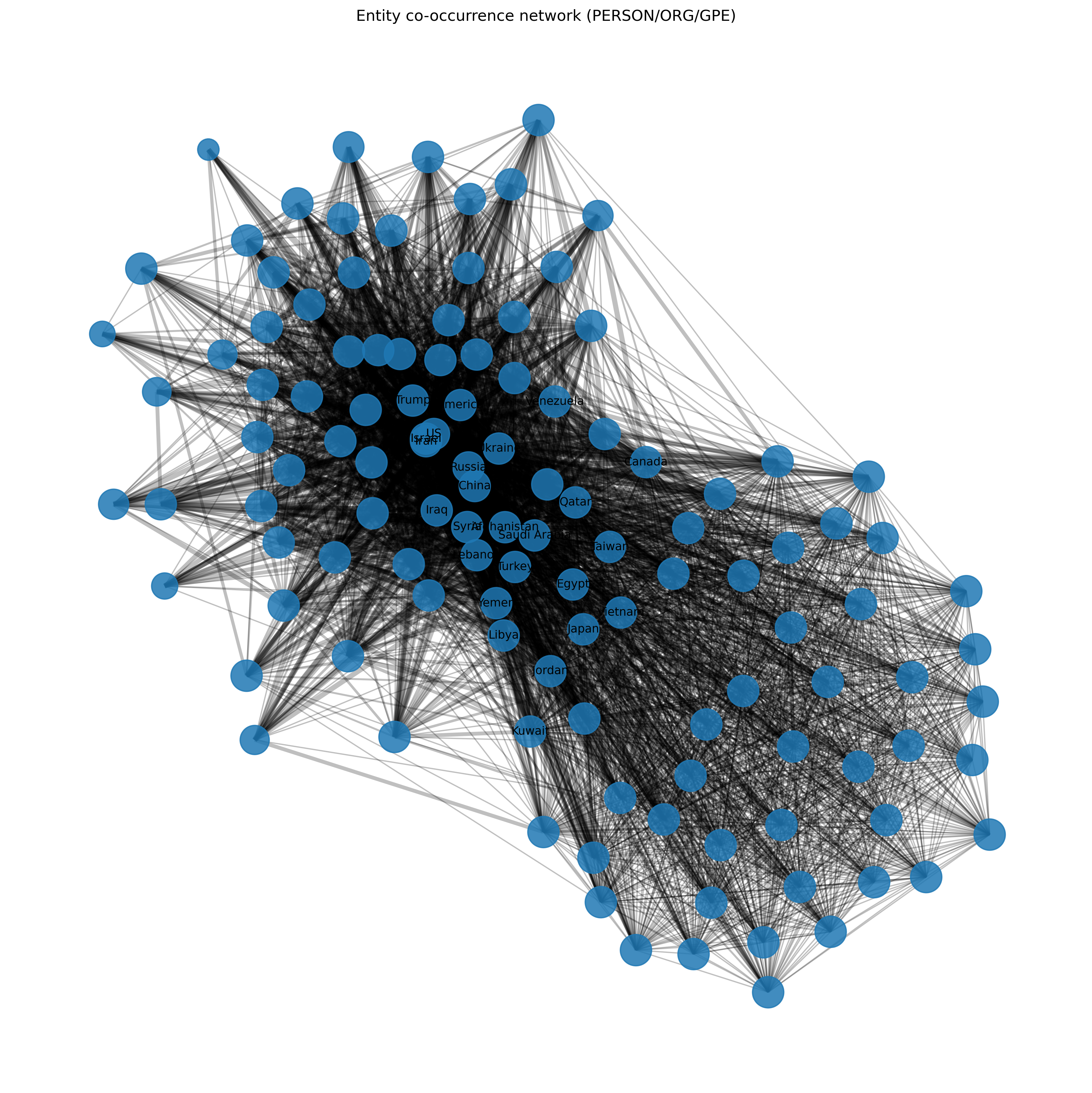}
\caption{Entity co-occurrence network derived from escalation-related discourse. Nodes represent geopolitical actors and locations, while edges represent co-occurrence relationships within messages.}
\label{fig:entity_network_rt}
\end{figure}

To examine the structural context of escalation discourse, we constructed an entity co-occurrence network based on named entities extracted from Telegram and Reddit messages. 
Figure~\ref{fig:entity_network_rt} shows strong connectivity among key geopolitical actors including Iran, the United States, Russia, Israel, and regional actors.

This network structure demonstrates that escalation narratives are embedded within broader geopolitical relationships, reflecting interconnected diplomatic, military, and strategic contexts.

\subsection{Summary}

Overall, the results demonstrate that the proposed framework successfully detects real-time narrative escalation across heterogeneous online platforms. Increases in message volume, escalation-related keyword activity, and escalation index values align with documented geopolitical developments. These findings confirm that cross-platform discourse analysis provides a reliable and interpretable indicator of emerging geopolitical escalation and evolving narrative dynamics.

\section{Related Work}

Social media platforms have become central arenas for political communication, enabling rapid dissemination of information, public debate, and narrative contestation. Prior research has demonstrated that online platforms significantly influence political discourse, public opinion formation, and narrative framing during political crises and geopolitical events \cite{antonakaki2017turbulence, jufri2023socialmedia}. These platforms facilitate both broadcast-style information dissemination and interactive discussion, contributing to the emergence of competing narratives and polarized viewpoints.

\subsection{Topic Modeling for Political and Social Media Analysis}

Topic modeling has emerged as a widely used approach for identifying latent thematic structures in large-scale textual datasets. Foundational methods such as Latent Dirichlet Allocation (LDA) and Non-negative Matrix Factorization (NMF) enable the identification of coherent topics and evolving discourse patterns without requiring predefined categories \cite{blei2003latent, grimmer2013text, roberts2014structural}. These techniques have been successfully applied to political speeches, social media datasets, and large textual corpora to identify key sociopolitical themes and discourse trends.

Recent research has extended topic modeling to geopolitical and crisis-related social media data, demonstrating its effectiveness in identifying protest narratives, conflict framing, and escalation-related discourse \cite{antonakaki2025crossplatform, lamprou2024crisis}. Topic modeling has also been applied to Telegram and Reddit data to identify extremist narratives, conspiracy theories, and coordinated geopolitical messaging \cite{steffen2025telegram, impicciche2025crossplatform}. These studies demonstrate that topic modeling provides an effective tool for identifying narrative structures and tracking their evolution across platforms.

\subsection{Sentiment Analysis and Polarization in Political Discourse}

Sentiment analysis provides a quantitative measure of emotional tone and affective framing in political discourse. Prior work has shown that sentiment analysis can identify shifts in public opinion, detect emotionally charged narratives, and characterize political polarization across social media platforms \cite{shevtsov2023tweets, noor2024canadianelection}. Emotional framing plays a critical role in shaping political narratives, influencing user perception and engagement.

Cross-platform sentiment analysis has revealed significant differences in emotional expression across platforms due to differences in communication structure, moderation policies, and user communities \cite{antonakaki2025crossplatform, lamprou2024crisis}. Polarization metrics such as Jensen--Shannon divergence and lexicon-based stance analysis have been used to quantify ideological divergence and identify polarized discourse communities \cite{ding2023semantic, hamdi2022extremism}. These approaches provide insight into how geopolitical narratives are emotionally framed and contested across different platforms.

\subsection{Cross-Platform Analysis of Political and Geopolitical Discourse}

Cross-platform analysis has become increasingly important for understanding how narratives propagate across heterogeneous online environments. Platforms such as Telegram and Reddit exhibit distinct communication affordances: Telegram primarily functions as a broadcast-oriented platform used for rapid information dissemination, while Reddit enables participatory discussion and community-driven interpretation \cite{willaert2023telegram, li2026metoo}. These structural differences influence narrative formation, emotional framing, and escalation dynamics.

Recent studies have examined cross-platform narrative propagation and coordination across Telegram, Reddit, and other platforms. Cross-platform topic modeling has been used to identify coordinated geopolitical narratives and ideological discourse across Telegram and Twitter \cite{kloo2024nazi}. Similarly, studies of coordinated inauthentic activity have demonstrated how narratives propagate across multiple platforms through amplification and reinforcement mechanisms \cite{cinus2025coordinated}. Cross-platform analyses have also shown that discourse responses to geopolitical events exhibit platform-specific dynamics, reflecting differences in audience, moderation, and communication structure \cite{buntain2023deplatforming, schulze2024affordance}.

These findings highlight the importance of analyzing multiple platforms simultaneously to capture the full structure and evolution of geopolitical discourse.

\subsection{Event Detection and Escalation Analysis in Social Media}

Social media platforms serve as real-time sensors of geopolitical developments, reflecting both immediate reactions and early indicators of emerging events. Prior work has demonstrated that keyword-based indicators, topic modeling, and sentiment dynamics can detect protest activity, political mobilization, and geopolitical escalation \cite{antonakaki2025israelhamas, lamprou2024crisis}. Temporal analysis of social media discourse has shown that narrative escalation frequently aligns with real-world geopolitical developments, while in some cases online discourse may precede documented events, reflecting early reporting or anticipation.

Event validation and correlation analysis have been used to assess the relationship between online discourse and real-world political developments, demonstrating that social media provides valuable signals of geopolitical escalation and narrative evolution.

\subsection{Research Gap and Contribution}

Although prior research has demonstrated the effectiveness of topic modeling, sentiment analysis, and cross-platform discourse analysis, several limitations remain. Many studies focus on single platforms, limiting the ability to capture cross-platform narrative propagation. Additionally, fewer studies integrate topic modeling, sentiment analysis, escalation detection, and event validation within a unified analytical framework.

To address this gap, the present study performs a comprehensive cross-platform analysis of Iran-related discourse across Telegram and Reddit. By integrating topic modeling, sentiment analysis, polarization measurement, and escalation detection with event correlation analysis, this work provides a unified framework for identifying and validating real-time narrative escalation across heterogeneous social media platforms. This approach enables systematic characterization of geopolitical discourse dynamics and demonstrates the feasibility of using cross-platform analysis for real-time escalation detection.

\section{Conclusion and Future Work}

This paper presented a comprehensive cross-platform analysis of Iran-related discourse on Telegram and Reddit, combining multi-year historical data with a separately collected escalation evaluation subset. Using a unified analytical pipeline based on TF--IDF representations, NMF topic modeling, sentiment analysis, polarization measurement, and keyword-based escalation indicators, we systematically characterized thematic structure, emotional framing, and temporal narrative dynamics across heterogeneous platforms.

Our results demonstrate clear structural differences between platforms: Telegram primarily reflects broadcast-style information dissemination driven by media organizations, while Reddit captures participatory discourse and interpretive discussion among users. Despite these structural differences, escalation-related narratives exhibit strong temporal alignment across platforms, indicating coordinated amplification of geopolitical narratives.

Importantly, correlation and lag analysis reveal that escalation-related discourse does not always occur simultaneously with documented geopolitical events. Instead, narrative escalation often precedes or follows real-world developments, suggesting that online discourse may function as an early indicator, anticipatory signal, or amplification mechanism rather than a simple real-time reflection of offline events. These findings highlight the value of cross-platform discourse monitoring for detecting emerging geopolitical developments.

\subsection{Interpretation of Topic Network Interconnections}

Beyond identifying lexical similarity, topic neighborhood networks provide insight into how narrative themes become discursively coupled. For example, the strong connection between Topic 7.0 and Topic 18.0 reflects identity-centered discourse rather than immediate geopolitical escalation. During the detected co-activity window (June 2025 – February 2026), discussions of Persian cultural heritage, diaspora identity, naming traditions, and regime critique co-occur across topics. 

This interconnection illustrates how cultural identity discourse becomes intertwined with political commentary. Rather than representing a single discrete event, such edges capture prolonged periods of identity negotiation and transnational framing. In contrast, other clusters within the network (e.g., military or nuclear topics) exhibit tighter temporal spikes and stronger alignment with documented geopolitical events. 

These distinctions suggest that topic similarity networks capture both event-driven escalation and structurally persistent identity discourse, highlighting the multifaceted nature of geopolitical narrative formation across platforms.

The escalation detection framework introduced in this study provides a reproducible and interpretable approach for quantifying narrative escalation using keyword bundle indicators normalized relative to baseline activity. Validation using independently collected data demonstrates that escalation-related narratives increase significantly during periods of geopolitical tension, supporting the effectiveness of the proposed methodology.

This work contributes a unified framework integrating topic modeling, sentiment analysis, polarization measurement, escalation detection, and event correlation within a single cross-platform analytical pipeline. By combining historical baseline analysis with escalation-focused evaluation, the framework enables systematic identification and validation of narrative escalation patterns.

Several limitations motivate future work. Platform coverage depends on selected channels, subreddits, and search queries, and keyword-based escalation indicators may not capture implicit or indirect escalation framing. Future research will extend this approach using embedding-based escalation classifiers, expand coverage to additional platforms, and evaluate escalation detection across diverse geopolitical contexts. Releasing datasets and analysis code will further enable reproducibility and comparative cross-platform studies.

Overall, the results demonstrate that cross-platform discourse analysis provides a reliable and interpretable method for detecting, characterizing, and monitoring geopolitical narrative escalation in online environments.
These findings suggest that large-scale cross-platform discourse monitoring can serve as a complementary signal for early detection of geopolitical escalation.

\section{Acknowledgments}

This research was supported by MEDIATE project (101074075 GA) funded by the European Union.

\bibliographystyle{acm}
\bibliography{main} 

\end{document}